\documentclass[useAMS,usenatbib]{mn2e}

\include{epsf}
\newcommand{\be}{\begin{equation}}
\newcommand{\ee}{\end{equation}}
\newcommand{\ba}{\begin{eqnarray}}
\newcommand{\ea}{\end{eqnarray}}

\def\simless{\mathbin{\lower 3pt\hbox
   {$\rlap{\raise 4pt\hbox{$\char'074$}}\mathchar"7218$}}}
\def\simgreat{\mathbin{\lower 3pt\hbox
   {$\rlap{\raise 4pt\hbox{$\char'076$}}\mathchar"7218$}}}
\title[SHADES Paper VII]
{The SCUBA HAlf Degree Extragalactic Survey (SHADES) -- VII. Optical/IR
photometry and stellar masses of sub-millimeter galaxies} 
\author[S. Dye et al.]{ 
\parbox[t]{\textwidth}{
S. Dye$^1$\thanks{E-mail: s.dye@astro.cf.ac.uk}, 
S. A. Eales$^1$,
I. Aretxaga$^2$, 
S. Serjeant$^3$, 
J. S. Dunlop$^4$, 
T. S. R. Babbedge$^5$,
S. C. Chapman$^6$,
M. Cirasuolo$^4$, 
D. L. Clements$^5$, 
K. E. K. Coppin$^7$,
L. Dunne$^8$,
E. Egami$^9$,  
D. Farrah$^{10}$,
R. J. Ivison$^4$,
E. van Kampen$^{11}$,
A. Pope$^{12}$,
R. Priddey$^{13}$,
G. H. Rieke$^9$,
A. M. Schael$^4$, 
D. Scott$^{12}$,
C. Simpson$^{14}$,
T. Takagi$^{15}$,  
T. Takata$^{16}$,
M. Vaccari$^{17}$
}\\ \\ 
$^1$Cardiff University, School of Physics \& Astronomy, Queens Buildings,
The Parade, Cardiff, CF24 3AA, U.K. \\
$^2$Instituto Nacional de Astrof\'{i}sica, \'{O}ptica y Electr\'{o}nica
(INAOE), Aptdo. Postal 51 y 216, 72000 Puebla, Pue., Mexico \\
$^3$Astrophysics Group, Department of Physics and Astronomy, 
The Open University, Milton Keynes, MK7 6AA, U.K. \\
$^4$SUPA\footnote{Scottish Universities Physics Alliance},
Institute for Astronomy, University of Edinburgh, Royal Observatory,
Edinburgh, EH9 3HJ, U.K. \\
$^5$Astrophysics Group, Blackett Laboratory, Imperial College, 
Prince Consort Road, London SW7 2BW, U.K.\\
$^6$Institute of Astronomy, University of Cambridge,
Madingley Road, Cambridge, CB3 0HA, U.K.\\
$^7$Institute for Computational Cosmology, Department of Physics,
Durham University, Durham, DH1 3LE, U.K.\\
$^8$The School of Physics \& Astronomy, University of Nottingham
University Park, Nottingham, NG7 2RD, U.K. \\
$^9$Steward Observatory, University of Arizona, 933 N. Cherry Avenue, 
Tuscon, AZ 85721, U.S.A. \\
$^{10}$Department of Astronomy, 610 Space Sciences Building, 
Cornell University, Ithaca, NY 14853, U.S.A \\
$^{11}$Institute for Astro- and Particle Physics, University of 
Innsbruck, Technikerstrasse 25, A-6020 Innsbruck, Austria\\
$^{12}$Department of Physics \& Astronomy, 
University of British Columbia, Vancouver, BC V6T 1Z1, Canada\\
$^{13}$Centre for Astrophysics Research, University of Hertfordshire,
Hatfield, AL10 9AB, U.K. \\
$^{14}$Astrophysics Research Institute, Liverpool John Moores University, 
Egerton Wharf, Birkenhead, CH41 1LD, U.K. \\
$^{15}$Institute of Space and Astronautical Science, Yoshinodai 3-1-1,
Sagamihara, Kanagawa 229 8510, Japan \\
$^{16}$National Astronomical Observatory of Japan,
650 North A'ohoku Place, Hilo, HI 96720, U.S.A.\\
$^{17}$Department of Astronomy, Vicolo dell'Osservatorio 3, 
University of Padova, I-35122 Padova, Italy \\
}

\begin{document}

\date{Document in prep.}

\pagerange{\pageref{firstpage}--\pageref{lastpage}} \pubyear{2006}

\maketitle

\label{firstpage}

\begin{abstract}

We present estimates of the photometric redshifts, stellar masses and
star formation histories of sources in the SCUBA HAlf Degree
Extragalactic Survey (SHADES). This paper describes the 60 SCUBA
sources detected in the Lockman Hole covering an area of $\sim 320$
arcmin$^2$.  Using photometry spanning the $B$ band to 8$\mu$m, we
find that the average SCUBA source forms a significant fraction of its
stars in an early period of star formation and that most of the
remainder forms in a shorter more intense burst around the redshift it
is observed. This trend does not vary significantly with source
redshift but the exact ratio of early to late mass is quite sensitive
to the way extinction is treated in the modelling. However, the
sources show a clear increase in stellar mass with redshift,
consistent with downsizing.  In terms of SED types, only two out of
the 51 sources we have obtained photometric redshifts for are best fit
by a quasar-like spectral energy distribution, with approximately 80
per cent of the sources being best fit with late-type spectra (Sc, Im
and starburst).  By including photometry at 850$\mu$m, we conclude
that the average SCUBA source is forming stars at a rate somewhere
between 6 and 30 times the rate implied from the rest-frame optical in
a dust obscured burst and that this burst creates 15-65 per cent of
the total stellar mass. Using a simplistic calculation, we estimate
from the average star formation history that between one in five and
one in 15 bright (${\rm L}_* +2 {\rm mag}<{\rm L_{optical}}<{\rm L}_*
-1 {\rm mag}$) galaxies in the field over the interval $0<z<3$ will at
some point in their lifetime experience a similar energetic dusty
burst of star formation. Finally, we compute the evolution of the star
formation rate density and find it peaks around $z \sim 2$.

\end{abstract}

\begin{keywords}
submillimetre - surveys - cosmology: observations - galaxies:
evolution - galaxies: formation - galaxies: high-redshift - infrared:
galaxies
\end{keywords}

\section{Introduction}
\label{sec_intro}

The SCUBA HAlf Degree Extragalactic Survey (SHADES) is a wide area
extragalactic sub-millimetre (submm) survey conducted with SCUBA
\citep[the Submillimetre Common User Bolometer Array;][]{holland99}. 
The motivation for SHADES is discussed at length in the survey
definition paper by \citet{mortier05}. The survey comprises two
separate fields of approximately equal area, one in the Lockman Hole
and one in the Subaru/XMM-Newton Deep Field (SXDF).  Up to the time of
decommissioning of SCUBA in 2005, SHADES had acquired approximately
40 per cent of the target area to the proposed depth of 2mJy, culminating in
the detection of a total of 120 robust SCUBA sources over $\sim 650$
arcmin$^2$ \citep{coppin06}.

SHADES satisfies a long awaited demand for a large, homogeneous sample
of submm sources with multi-wavelength follow-up data. Since their
detection in the first deep submm surveys
\citep{smail97,hughes98,barger98}, much has been learnt about the
dusty high-redshift sources revealed by SCUBA. However, several major
questions regarding this elusive population remain unanswered.

Arguably the most important question is the relationship between SCUBA
sources and present-day galaxies.  Several clues point toward a strong
link with massive low-redshift ellipticals, such as their similar
comoving number densities \citep{scott02,dunne03}, clustering
properties \citep[e.g.,][]{almaini03,blain04} and their typically very
high star formation rates which enable the rapid formation of a large
stellar system.  If this link is valid, then an immediate question
that arises is at what stage of this transformation do we observe the
object as a SCUBA galaxy? It is possible that there is more than one
answer if multiple routes exist to the same type of final massive
elliptical.  For example, the average elliptical's entire stellar
population may either form in a single large burst or in a series of
smaller bursts triggered by mergers.

Crucial evidence can be provided by the star formation history of the
average SCUBA source.  However, in order to establish a star formation
history, several ingredients are needed: 1) the source must be
identified from its somewhat imprecise SCUBA position, 2)
multi-wavelength data (ideally covering optical to submm) must be
acquired 3) the source's redshift must be known. Deep radio surveys
\citep[e.g.,][]{ivison02} detect somewhere between a half and
three-quarters of SCUBA sources to give precise positional
information.  In addition, SCUBA sources can be efficiently identified
with relatively short exposures using the Spitzer Space Telescope
({\sl Spitzer}) as demonstrated by several authors
\citep[e.g.][]{huang04,egami04}.  Once accurate positions have been
obtained, optical spectroscopy can then be carried out.

Unfortunately, this procedure has several selection effects. The
requirement that a source be detected at radio wavelengths can lead to
a lack of sources at high redshifts ($z \simgreat 3$) where the radio
flux falls below the detection limit \citep[see for
example][]{chapman05}. There is also a paucity of sources at redshifts
where no bright spectral features fall within the optical waveband,
particularly over the redshift interval $1.2<z<1.9$ (the `redshift
desert'). Photometric redshifts, albeit less precise, do not suffer
from the latter selection effect.  Another advantage is that
contaminating flux from near-neighbours and blended sources (as an
appreciable fraction of SCUBA sources appear to be) is more readily
quantified in image data unlike spectroscopic data where small
uncertainties in the slit placement can lead to ambiguities in
deblending.

In this paper, we make use of photometric redshifts to investigate the
photometric properties, stellar masses and star formation histories of
SCUBA sources in SHADES. Counterparts are identified through either
deep radio data or {\sl Spitzer} 24$\mu$m images.  Having the option
of a 24$\mu$m identification means that the SCUBA sources considered
in this work are not entirely subject to the strong radio selection
function (although it may be contended that the process of identifying
counterparts in the radio is more physically motivated).

This paper is seventh in a series of papers arising from SHADES.
Paper I by \citet{mortier05} describes the science goals, motivation
and strategy. Paper II \citep{coppin06} presents the maps, catalogues
and source counts and describes the data reduction. Paper III
\citep{ivison07} details the radio follow-up of the SHADES areas and
identifies the radio and 24$\mu$m {\sl Spitzer} counterparts to the
SHADES sources.  Paper IV \citep{aretxaga07} derives photometric
redshifts of the SHADES sources using radio, submm and far infrared
(far-IR) data.  Paper V \citep{takagi07} concerns the submm properties
of near-IR selected galaxies in the SXDF. Paper VI \citep{coppin07}
presents 350$\mu$m observations of of a subset of SHADES sources.
Papers VII (this paper) and VIII \citep{clements07} form a pair split
by survey area: paper VII considers photometric redshifts, stellar
masses and star formation histories of sources in the Lockman Hole
whereas paper VIII is concerned with photometric redshifts in the
SXDF.  Papers VII and VIII are divided primarily due to data
propriety, but also because of different optical and near-IR coverage
between the two areas.  In terms of forthcoming papers, paper IX
\citep{serjeant07} will investigate stacking of the SHADES data to
statistically determine properties of the very faint sources hidden in
the noise. Paper X \citep{vankampen07} will study the clustering of
SHADES sources.  Finally, several further SHADES publications are
anticipated concerning 1.1mm data acquired with the AzTEC instrument
on the James Clerk Maxwell Telescope.

The layout of this paper is as follows. Section \ref{sec_data}
outlines the acquisition and reduction of data used in this
work. Section \ref{sec_photz} discusses the results of our photometric
redshift analysis and properties of the SEDs. In Section
\ref{sec_stellar_mass}, we describe our method of obtaining stellar
masses together with the resulting masses, star formation rates and
evolution of the star formation rate density.  We conclude the main
paper sections with a summary and discussion in Section
\ref{sec_summary}. We provide three appendices: Appendix
\ref{sec_app_seds} contains multi-wavelength postage stamp images for
each of our sources as well as their best fit SEDs, Appendix
\ref{sec_app_source_notes} provides descriptions for a selection of
noteworthy sources and Appendix \ref{sec_app_photom_data} lists
the photometry for all sources.

Throughout this paper, we assume the following cosmology: 
H$_0$=70kms$^{-1}$Mpc$^{-1}$, $\Omega_{\rm m}=0.3$,
$\Omega_\Lambda=0.7$.

\section{Data}
\label{sec_data}

\subsection{SHADES catalogue \& counterparts}
\label{sec_shades_cat}

The sources investigated in this paper were extracted from 850$\mu$m
SCUBA observations of the Lockman Hole centred on RA=$10^{\rm
h}52^{\rm m}26.7^{\rm s}$, Dec=$57^{\circ}24'12.6''$ (J2000). The
Lockman Hole data cover an area of $\sim 320\, {\rm arcmin}^2$ to an
RMS noise level of $\sim 2$mJy.  To ensure a robust list of SCUBA
sources, the map reduction and catalogue generation was carried out by
four independent groups within the SHADES consortium. Only sources
with a signal to noise ratio of $\geq 3$ (before deboosting) in at
least two reductions were retained. This left a total of 60 sources in
the Lockman Hole field with a probability of $< 5$ per cent of being
spurious. For more specific details on the reduction of the SCUBA map
and source extraction, the reader is referred to
\citet{coppin06}.

Follow-up 1.4GHz imaging with the Very Large Array and 24$\mu$m
imaging with {\sl Spitzer} are described in \citet{ivison07}.
Potential radio and 24$\mu$m counterparts to the SCUBA sources were
searched for within a radius of $8''$ and the significance of each
match quantified using the method of \citet{downes86}.  This method
gives the probability, $P$, of a counterpart being associated with the
SCUBA position by chance based on the separation and the number
counts. In this way, a counterpart either in the radio or at 24$\mu$m
is defined as being robust if its distance from the SCUBA position is
less than $8''$ and $P\leq 0.05$.  Non-robust identifications are
defined as those with $P>0.05$ within $8''$ or those that lie within
an extended search radius $8''<r<12.5''$ in the radio or $8''<r<15''$
at 24$\mu$m.

\citet{ivison07} lists several counterparts for many of the SHADES
sources. In the present work, we base our analysis on the single, most
likely counterpart for each source. We define these `primary
counterparts' as those with the lowest value of $P$ in the radio or at
24$\mu$m (often both). Although not included in any of our later
analyses, for completeness, we also compute photometric redshifts for
secondary counterparts. A secondary counterpart is defined as having a
robust 24$\mu$m and/or robust radio identification but with a
numerically higher $P$ than the primary. These are listed in Table
\ref{tab_sources} alongside the primary counterparts. In all cases,
the co-ordinates listed in Table \ref{tab_sources} are either the
radio or 24$\mu$m co-ordinates given in \citet{ivison07} depending on
which identification has the lowest $P$. (There is one exception:
Lock850.036 for which we give the SCUBA 850$\mu$m centroid as this
source has no radio or 24$\mu$m counterparts).

\subsection{Optical and near-IR photometry}

Our optical images in $B$, $R$, $I$ and $z$ were obtained with
SuprimeCam \citep{miyazaki02} on the Subaru telescope in January
2006. The images fully cover the Lockman area observed by SCUBA and
reach a $5\sigma$ point source sensitivity of 26.8, 25.8, 25.7 and
25.0 mag (AB) in $B$, $R$, $I$ and $z$ respectively as measured in a
$3''$ diameter aperture. Total exposure times for $B$, $R$, $I$ and
$z$ are respectively 7200s, 3360s, 4730s and 4800s. The seeing in the
images varies from $0.66''$ to $0.84''$ between bands with a mean of
$0.76''$.

Our $K$ band image was obtained with the Wide Field Camera (WFCAM) on
the United Kingdom Infra-red Telescope (UKIRT). The data were taken
as part of the Deep Extra-galactic Survey, one of the five projects
comprising the UKIRT Deep Infra-red Sky Survey
\citep[UKIDSS][]{lawrence06}. The image is a mosaic of four separate
quadrants observed in array number 1 \citep[see][for further
details]{dye06}. The exposure time of the four quadrants varies
between 9180s and 11460s, reaching an average $5\sigma$ point source
sensitivity of 22.9 mag (AB) \citep{warren07}.

We used {\sl SExtractor} \citep{bertin96} to extract sources from the
optical and $K$ band data. Only objects with five or more
interconnecting pixels lying above a threshold signal-to-noise of
2$\sigma$ were extracted. Fluxes were computed within a $3''$ diameter
aperture.

\subsection{{\sl Spitzer} photometry}

{\sl Spitzer}'s Infra-red Array Camera \citep[IRAC;][]{fazio04} was
used to obtain images at wavelengths 3.6, 4.5, 5.8 and 8$\mu$m. The
integration time for each image was 500s, reaching $5\sigma$ point
source detection limits of 1.3, 2.7, 18 and 22$\mu$Jy for 3.6, 4.5,
5.8 and 8$\mu$m respectively.  The data were reduced using the {\sl
Spitzer} Science Centre's (SSC) pipeline \citep{gordon05}. 

The pipeline produces the total flux of each source. To match the
total flux of a given source in each of the IRAC bands to the optical
and near-IR $3''$ photometry accounting for the different PSFs, we
adopted the following procedure: 1) Fit a 2D Gaussian to the source in
the IRAC image, 2) Scale the fitted Gaussian to the size it would have
been if it had been observed with the SuprimeCam PSF, 3) Compute the
flux within a $3''$ diameter aperture centred on the scaled Gaussian
using the total flux output by the IRAC pipeline.  The average of all
corrections across all bands and sources\footnote{Since the correction
applies to the total flux which is practically insensitive to the PSF
and since sources are effectively brought to the same seeing in each
IRAC waveband, the scatter in the correction between wavebands for a
given source is primarily due to the image noise and wavelength
dependent source morphology, not the varying PSF between wavebands.}
was +0.34 mag with a $1\sigma$ scatter of 0.15 mag.  To account for
possible systematics introduced by this scheme, we added an error of
0.15 mag in quadrature to the error computed by the SSC pipeline for
all IRAC photometry.

In addition to IRAC imaging of the Lockman Hole, data were taken using
the Multi-band Photometer for {\sl Spitzer} \citep[MIPS;][]{rieke04}.
Although we do not use MIPS fluxes directly in the present work, our
identification of counterparts to the SCUBA sources relies on 24$\mu$m
MIPS detections \citep[see previous section and also][]{ivison07}.

\subsection{Cross matching procedure}

To obtain the multi-wavelength list for all the SHADES counterparts,
the sources on the different images were position coincidence matched
using a tolerance of $1''$. All matches were carefully verified by eye
to correct for obvious mismatches, spurious detections and blended
photometry. In cases where blended photometry was identified, we
adjusted the the deblending threshold and cleaning parameter in
{\sl SExtractor} on a source by source basis until deblended photometry
was obtained.

\section{Photometric redshifts and Spectra}
\label{sec_photz}

\subsection{Method of determination}
\label{sec_photz_method}

We obtained photometric redshifts for the SHADES sources by applying
the {\sl HyperZ} redshift code \citep{bolzonella00} to our nine band
photometry ($B$, $R$, $I$, $z$, $K$, 3.6$\mu$m, 4.5$\mu$m, 5.8$\mu$m
and 8$\mu$m) described in the previous section. The template galaxy
spectral energy distributions (SEDs) packaged with {\sl HyperZ} are
based on the local SEDs measured by \citet{coleman80}, with an
extrapolation into the infrared using the results of spectral
synthesis models.  The lack of any empirical basis for these templates
at wavelengths $>1\mu$m is clearly unsatisfactory since our photometry
includes both near-infrared and mid-infrared measurements. Therefore,
we constructed our own set of templates. \citet{mannucci01} list
empirical SEDs over the wavelength range 0.1$\mu$m to 2.4$\mu$m for
the Hubble types E, S0, Sa, Sb and Sc. We extended these out to
10$\mu$m using the average SEDs for disk galaxies and elliptical
galaxies listed in Table 3 of \citet{lu03}.

One disadvantage of these templates is the lack of a template for
an irregular galaxy, and we therefore retained the {\sl HyperZ} Im
template, extending this into the mid-infrared using
the average disk-galaxy SED from \citet{lu03}. In terms of the
variation in SED shape, particularly the average slope between
$0.2\mu$m$<\lambda<2\mu$m, the jump from the Sc template to the Im
template is larger than the progression through the earlier type
SEDs. Therefore, we introduced a seventh intermediate template with a
composition of 60 per cent Sc and 40 per cent Im to bridge this jump. 

Finally, we introduced a starburst (SB) and quasar (QSO) template.
For the SB template, we extended the spectrum of \citet{kinney96} to
longer and shorter wavelengths using a \citet{bruzual03} spectrum
corresponding to a 0.1Gyr duration starburst as observed at the end of
that time. Since this extension lacks the dusty spectral features seen
in the mid-IR, we added these from the spectra of \citet{lu03}. The
QSO template was taken from \citet{brotherton01} and extended longward
of 7500 \AA using the QSO spectrum of \citet{elvis94}.  This is an
average SED and therefore does not cater for a variable power-law
slope.  Similarly, we chose not to include templates containing a mix
of QSO and starburst activity.  Our preference is for a more clear-cut
approach: The QSO template that we use is sufficiently different from
the other templates to be well suited for picking out obvious QSO
candidates.

\begin{figure}
\epsfxsize=80mm
{\hfill
\epsfbox{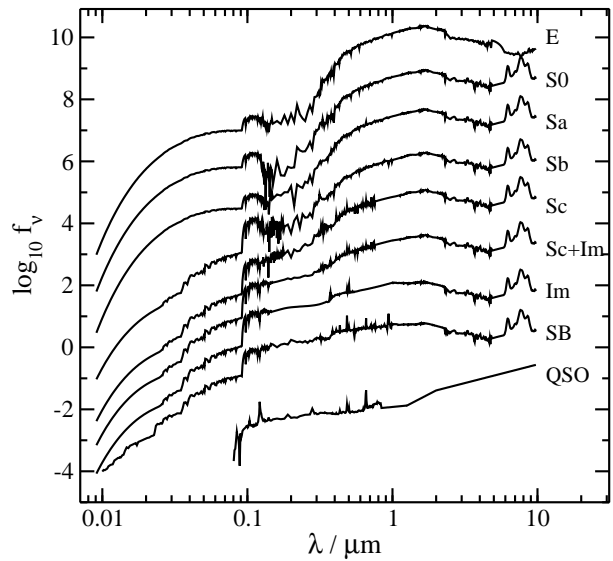}
\hfill}
\caption{Template SEDs used with {\sl HyperZ} for determination
of the photometric redshifts. The template Sc+Im is a hybrid SED
composed of 60 per cent Sc and 40 per cent Im to bridge the jump
between the Sc and Im types (see text). SED templates are referred to
throughout this paper by their number; 1=E, 2=S0, 3=Sa, 4=Sb, 5=Sc,
6=Sc+Im, 7=Im, 8=Starburst, 9=QSO.}
\label{sed_templates}
\end{figure}

The nine SED templates are plotted in Figure
\ref{sed_templates}. Throughout this paper, we refer to SED templates
by their number; 1=E, 2=S0, 3=Sa, 4=Sb, 5=Sc, 6=Sc+Im, 7=Im,
8=Starburst, 9=QSO.

To determine the redshifts, we allowed {\sl HyperZ} to search over the
range $0<z<6$. We used a Calzetti reddening law to account for dust
extinction, allowing the rest-frame $V$-band attenuation, $A_V$, to vary
from 0 to 5 mag in steps of 0.1 mag (requiring re-compilation of {\sl
HyperZ} with a larger array). No cut-off was imposed on the absolute $K$
band magnitude (AB) computed by {\sl HyperZ} (see Section
\ref{sec_stellar_mass}) although we imposed a minimum magnitude error
of 0.05 mag for all filters to account for zero point errors.  We
configured {\sl HyperZ} to treat non-detected sources as having zero
flux with a $1\sigma$ error equal to the flux sensitivity of the
corresponding filter. (We also tried a more stringent configuration
whereby the error is set to half the sensitivity with negligible
effect on the results). All filter transmission profiles were modified
where appropriate to account for the wavelength-dependent quantum
efficiency of the detector, the mirror reflectivity and atmospheric
attenuation.

\begin{figure}
\epsfxsize=80mm
{\hfill
\epsfbox{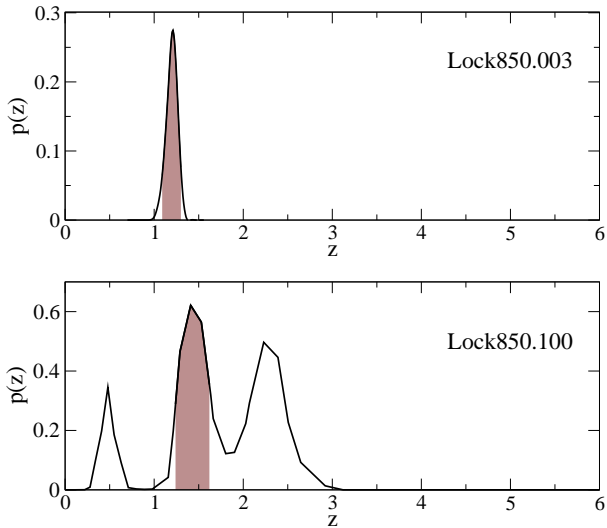}
\hfill}
\caption{Two extreme examples of the photometric redshift probability
distribution of the SHADES sources studied in this paper.  The top
panel corresponds to the source Lock850.003 which has a very robust
photometric redshift. Conversely, the bottom panel corresponds to
Lock850.100 which has three almost equally likely photometric
redshifts. The grey shaded area indicates the 90\% confidence range in
primary $z$ as given in Table \ref{tab_sources}.}
\label{p_of_z_extremes}
\end{figure}

We note that we include the quantity p($z$) in Table \ref{tab_sources}
and its corresponding p($z2$) for secondary redshifts. This quantity
is the integral of the area under the peak of the redshift probability
distribution normalised by the total area within $0<z<6$
for a given source. The redshift probability distribution
is computed from the $\chi^2$ curve output by {\sl HyperZ}. The most
likely redshift (i.e., lowest $\chi^2$) is taken as the primary
solution, not the peak with the largest enveloped area. Figure
\ref{p_of_z_extremes} shows p($z$) versus $z$ for two extreme
cases in Table \ref{tab_sources}. At one extreme is Lock850.003 where
a very robust photometric redshift with a small error and no secondary
solution is obtained and at the other extreme is Lock850.100 where three
almost equally likely solutions exist. We find that the majority of
sources have a single well defined peak.

\subsection{Photometric redshifts and SED results}

\subsubsection{Photometric redshifts}
\label{sec_photz_results}

\begin{figure}
\epsfxsize=80mm
{\hfill
\epsfbox{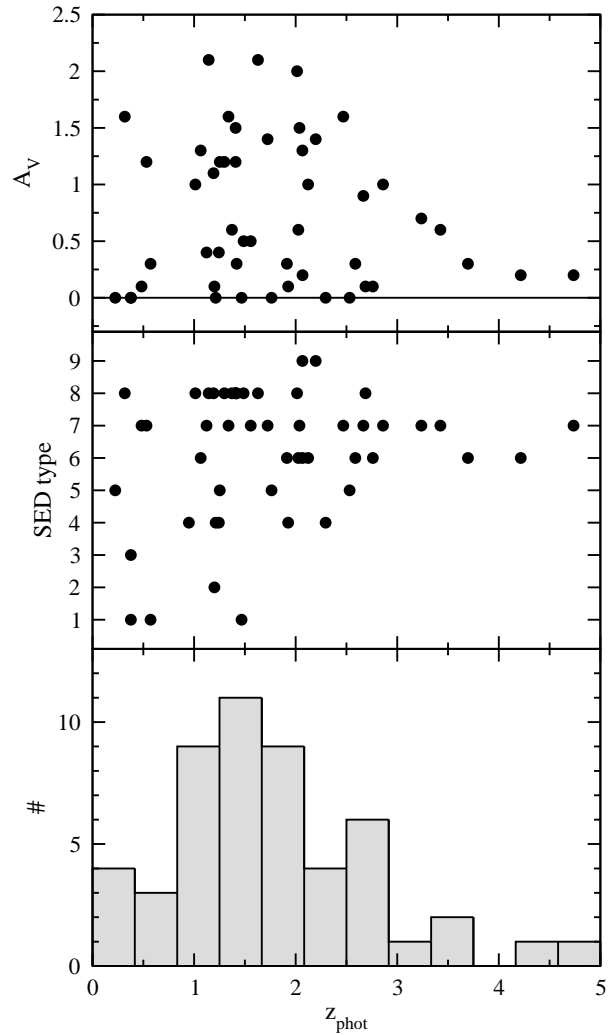}
\hfill}
\caption{{\em Bottom panel}: Histogram of photometric redshifts
computed for the 51 primary SHADES sources listed in Table
\ref{tab_sources}.  {\em Middle panel}: Variation of SED type with
redshift. {\em Top panel}: Variation of attenuation by dust with
redshift.}
\label{sed_properties}
\end{figure}

Table \ref{tab_sources} lists the photometric redshifts obtained for
51 primary counterparts (and 12 secondary counterparts) of the 60
SHADES sources and the lower plot in Figure \ref{sed_properties} shows
their distribution. The best fitting SEDs are shown in Figure
\ref{seds}. The median redshift of the sample is $z=1.52$ in close 
agreement with the median photometric redshift measured in the
SHADES SXDF field by \citet{clements07}. Redshifts for nine of the
sources could not be determined due to insufficient photometric points
(either because they are too faint and therefore not detected at
certain wavelengths -- we stipulate a minimum of three fluxes per
source to determine a redshift -- or because they are heavily blended
with a neighbouring source).

A simple measure of the appropriateness of our SED templates to the
photometry is the distribution of $\chi^2$. This analysis depends on
the accuracy of our photometric errors although for simple aperture
fluxes, these are reliably determined. The distribution for all our
SED fits is perfectly consistent with a $\chi^2$ distribution of 5
degrees of freedom. This is as it should be for nine photometric
constraints (including upper limits) and four fitted parameters ($z$,
$A_V$, SED normalisation and SED type) implying that our SED template
set provides a suitable match to the SCUBA sources.

We have compared the photometric redshifts output by {\sl HyperZ} with
three alternative determinations (see Table \ref{tab_z_compare} for a
summary). Firstly, still applying {\sl HyperZ} but using the synthetic
spectra of \citet{bruzual03} packaged with it, we found generally good
agreement although a slightly larger spread in the distribution of
$\chi^2$. Secondly, we used the {\sl BPZ} code of \citet{benitez00}
with our own SED templates.  Although this code does not allow for
attenuation by dust, it uses a Bayesian approach incorporating a
luminosity prior.  We set the prior to the luminosity function
obtained from the Hubble Deep Field North and found excellent
agreement with the results of {\sl HyperZ} when no dust attenuation
was included. However, the inclusion of dust attenuation in {\sl
HyperZ} gives significantly different redshifts for a large fraction
of the sources and values of $\chi^2$ that are a factor of two times
lower on average. As expected, this shows that dust has a significant
effect on the SEDs of the SCUBA sources (also reflected in the values
of $A_V$ seen in Table \ref{tab_sources}).

\begin{figure}
\epsfxsize=75mm
{\hfill
\epsfbox{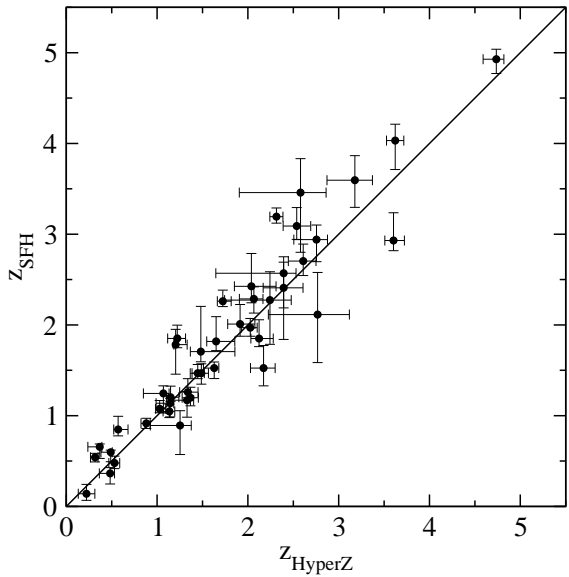}
\hfill}
\caption{Comparison of the photometric redshifts computed by
{\sl HyperZ} and those from the star formation history analysis of 
Section \ref{sec_stellar_mass_method} (1$\sigma$ error bars plotted).}
\label{hyperz_vs_my_z}
\end{figure}

In the third and final case, we compared the {\sl HyperZ} redshift
estimates with our own photometric redshifts obtained using the star
formation history analysis presented in Section
\ref{sec_stellar_mass_method}. Figure \ref{hyperz_vs_my_z} shows the
comparison. The agreement is very good with an average $|\Delta
z|/(1+z)$ of 0.09 (for comparison, the average $\sigma_z/(1+z)$ for
the {\sl HyperZ} redshifts is 0.06).  Since the two methods are
completely independent, with one using empirical SED template fitting
and the other synthetic spectra constructed from a best fit star
formation history, this supports the reliability of our redshifts.

\begin{figure}
\epsfxsize=75mm
{\hfill
\epsfbox{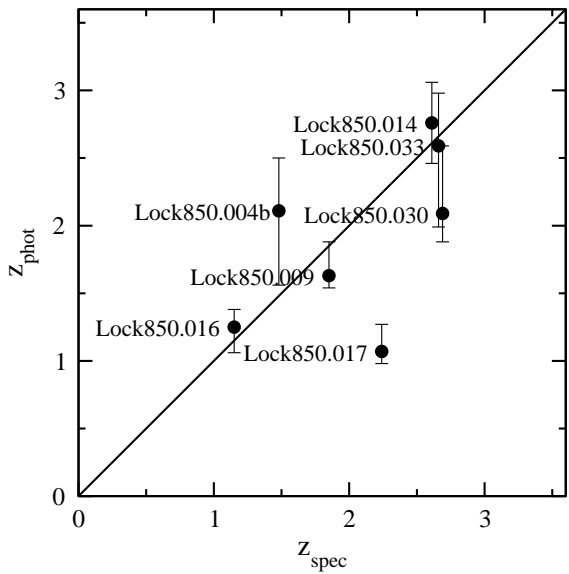}
\hfill}
\caption{Comparison of photometric redshifts and robust spectroscopic 
redshifts from \citet{ivison05} and \citet{chapman05}. Error bars show
the 90 per cent confidence range on the photometric redshifts. }
\label{zspec_vs_zphot}
\end{figure}

Unfortunately, there are very few SCUBA sources in the Lockman Hole
with robust spectroscopic redshifts against which we can compare our
photometric redshifts. Nevertheless, in Figure \ref{zspec_vs_zphot} we
show a comparison with the 7 that currently exist.  The only
irreconcilable discrepancy is Lock850.017 with $z_{\rm
phot}=1.06^{+0.24}_{-0.06}$ and $z_{\rm spec}=2.24$. The optical
counterpart to this source coincides exactly with the radio and
spectroscopic position, is not blended with any neighbours and is
detected with a very high signal to noise at all wavelengths. The
photometry is extremely well fit by the Im+Sc template with a reduced
$\chi^2$ of $0.28\pm0.10$ and a single sharp peak in the probability
distribution of the fit.  However, the optical morphology shows a
relatively diffuse tail extending $\sim 3''$ to the south-west of the
nucleus of this source. A possible cause of the discrepancy could
therefore be that there are two closely aligned objects at different
redshifts.  Our photometry is dominated by the compact nucleus which
could be a foreground source, whereas the spectroscopic redshift could
be based on emission lines from the more diffuse background
source.

\begin{table}
\centering
\begin{tabular}{ccc}
\hline
Photo-z type & $<|\Delta z|>$ & $<\sigma_z>$ \\
\hline
{\sl HyperZ} & $-$ & 0.10 \\
{\sl HyperZ}+BC & 0.34 & 0.14 \\
{\sl BPZ} & 0.21 & 0.31 \\
SFH & 0.18 & 0.16 \\
submm & 0.78 & 0.60 \\
\hline
\end{tabular}
\caption{Comparison of the different photometric redshifts (photo-zs)
considered in
Section \ref{sec_photz_results}. $<\sigma_z>$ is the median $1\sigma$
uncertainty in the photo-zs of all 60 SHADES sources and $<|\Delta
z|>$ is the median absolute difference between the photo-zs
listed in Table \ref{tab_sources} and the others 
discussed. The photo-z types listed are: 
{\sl HyperZ} -- the photo-zs in Table
\ref{tab_sources}, i.e., obtained using {\sl HyperZ} with the empirical
SED templates discussed in Section \ref{sec_photz_method};
{\sl HyperZ} + BC -- again using {\sl HyperZ} but with the
\citet{bruzual03} SED templates; 
{\sl BPZ} -- using the code by \citet{benitez00} with the emperical
SED templates of Section \ref{sec_photz_method};
SFH -- our own photo-zs from the star formation history 
analysis of Section \ref{sec_stellar_mass_method};
submm -- the photo-zs of \citet{aretxaga07}.}
\label{tab_z_compare}
\end{table}

We draw particular attention to the interesting source Lock850.041b.
The radio co-ordinates of this source correspond to the brightest but
slightly less favourable of two radio sources within $8''$ of the
SCUBA centroid. \citet{ivison05} were unable to measure a
spectroscopic redshift at the radio co-ordinates of Lock850.041b but
did measure a spectroscopic redshift of $z=0.69$ for an elliptical
galaxy lying $\sim 2''$ to the south-west of the radio
co-ordinates. We determine a photometric redshift of $z=2.22$ for the
radio source and find that the photometry is well fit by a QSO SED
template. Combined with the optical morphology which exhibits arc-like
structure (the arcs being a result of a relatively bright extended
host galaxy), these facts suggest that Lock850.041b is very likely a
strongly lensed QSO. This supports the claim of \citet{chapman02} that
the source is a plausible lens.

\begin{figure}
\epsfxsize=75mm
{\hfill
\epsfbox{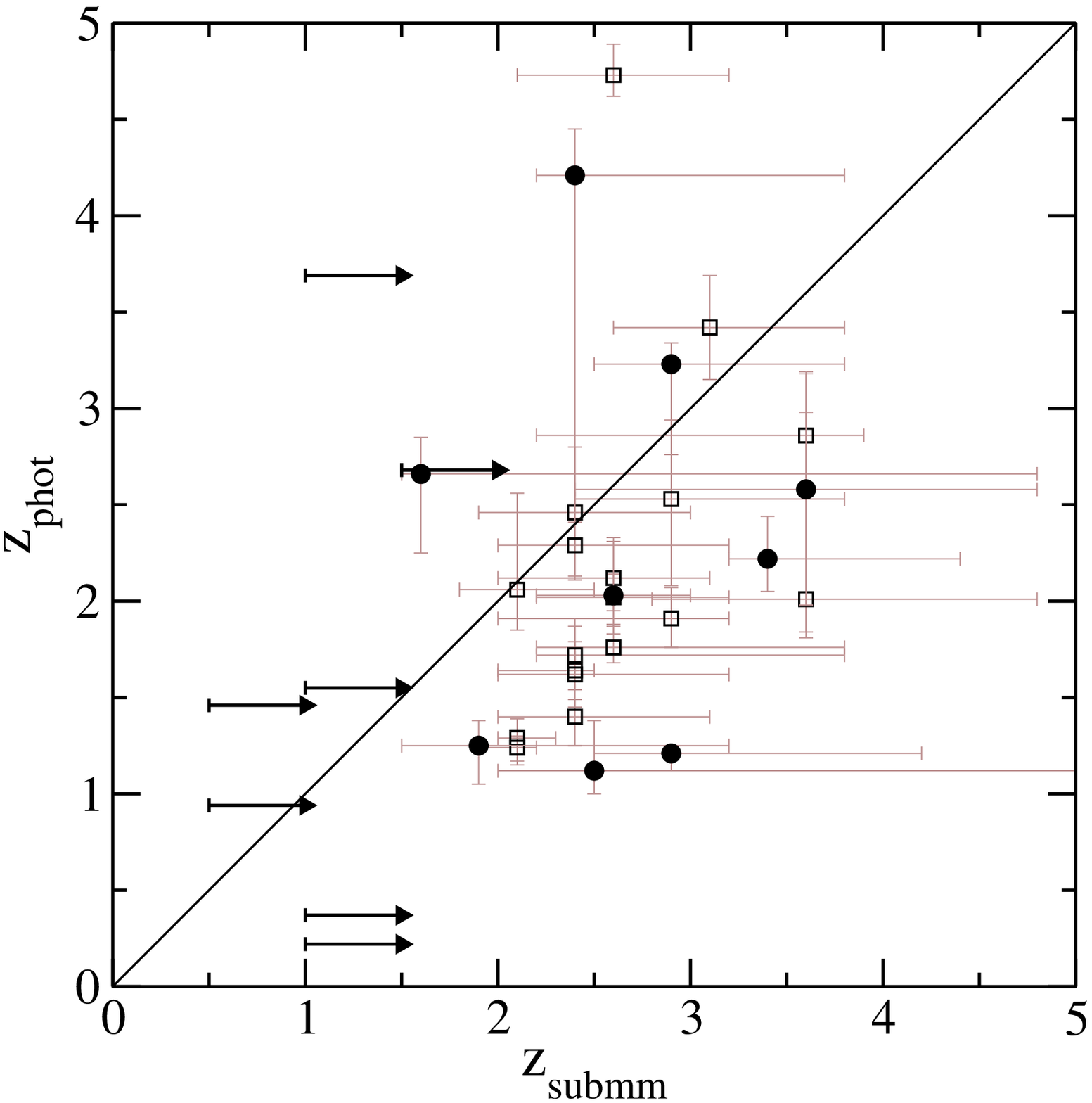}
\hfill}
\caption{Comparison of photometric redshifts computed in this work
($z_{\rm phot}$) with the photometric redshifts of \citet{aretxaga07}
derived using the 850$\mu$m and 1.4GHz spectral index ($z_{\rm
submm}$).  Open squares correspond to redshifts derived solely from
850$\mu$m and 1.4GHz photometry whereas filled circles correspond to
redshifts derived with additional far-IR/submm photometric
data. Both vertical and horizontal error bars denote the 90 per cent
confidence range. The tabbed ends of the arrows show the lower limits
set by \citet{aretxaga07}.}
\label{zphot_vs_zphot_submm}
\end{figure}

Figure \ref{zphot_vs_zphot_submm} compares our redshifts with the
photometric redshifts of \citet{aretxaga07} determined using the
850$\mu$m/1.4GHz spectral index (their `$z^{\rm MC}_{\rm phot}$'). We
find a large but not significant offset between both sets of redshifts
(not including sources with lower limits). Comparing the sources with
$z_{\rm submm}$ derived from more than two photometric points (the
filled circles in Figure \ref{zphot_vs_zphot_submm}), our redshifts
are lower by $\Delta z=0.37 \pm 0.39$ on average, where the error here
is the error on the mean. Extending this comparison to include all
sources in Figure \ref{zphot_vs_zphot_submm}, we find that our
redshifts are lower by an average of $\Delta z=0.42 \pm 0.35$ (or
alternatively, as given in Table \ref{tab_z_compare}, the median value
of $|\Delta z|$ is 0.78).

This offset manifests itself more strongly when comparing the peak of
our total redshift distribution at $z\simeq 1.5$ to that of
\citet{aretxaga07} which lies at $z\simeq 2.3$ \citep[][ also find
that the peak of their spectroscopic redshift distribution lies at
$z\simeq 2.3$]{chapman05}. The origin of this discrepancy is not
entirely clear, although the offset is reduced if the comparison is
limited to the redshift distributions of only those sources plotted in
Figure \ref{zphot_vs_zphot_submm} (i.e., those with redshifts
determined by both the optical and submm/radio methods).  This
suggests that part of the discrepancy is a selection effect. If this
is the case, then it is interesting to note how the two distributions
compare to the models of \citet{vankampen05}. A KS test shows that
our redshift distribution is consistent with the massive merger
model whereby all SCUBA galaxies are the result of a violent merger
of two galaxy sized halos. Conversely, the redshift distribution of
\citet{aretxaga07} is best fit by the phenomenological model, which
includes a mixture of starbursts due to mergers and quiescent star
formation. \citet{aretxaga07} note that their redshifts are even
better matched by the distribution predicted from the semi-analytical
model of \citet{silva05} which jointly describes the formation and
evolution of spheroids and QSOs. An implication of this might
therefore be that SCUBA sources more readily detected in the optical
are more likely to be the result of massive mergers, whereas a pure
submm selected sample is more likely to include sources with quiescent
star formation.

\subsubsection{SED properties}

Figure \ref{seds} in Appendix \ref{sec_app_seds} shows the best fit
SEDs to the SHADES sources for which we have been able to estimate
redshifts. The SED type and $A_V$ given in Table \ref{tab_sources} are
plotted in Figure \ref{sed_properties} as a function of
redshift. There are two distinct trends. Firstly, there is a clear
evolution in SED type, with late types being found predominantly at
higher redshifts and early types at lower redshifts. Secondly, sources
at $z>2.5$ have significantly lower attenuation than sources at
$z<2.5$: Dividing the top panel of Figure \ref{sed_properties} by the
line $A_V=1$, it can be seen that sources at $z<2.5$ are split
equally, whereas at $z>2.5$, 11 sources have an attenuation $A_V<1$
with none having $A_V>1$. We have verified that this is not a result
of the way {\sl HyperZ} deals with shorter waveband dropouts by
reproducing this trend independently of the configured method (see
Section \ref{sec_photz_method}).

Such a decline in attenuation at high redshifts is consistent with the
rate of dusty star formation having peaked at an earlier epoch (see
Section \ref{sec_sfr_evol}), although there is a strong selection
effect due to the sensitivity of the data. This decline is also
reflected in the declining number of SHADES sources found beyond
$z>2.5$. This is not entirely a result of the usual selection effects
since our redshifts are not subject to the optical spectroscopic
redshift desert discussed in Section \ref{sec_intro}, nor are they
totally reliant on there being a radio detection since the ID
procedure also relies on detections at 24$\mu$m.  A similar decline in
attenuation due to dust has recently been found by
\citet{buat07}. Their findings indicate a monotonic decrease in dust
attenuation in luminous infrared galaxies from $z=0$ out to $z=2$ for
a fixed star formation rate \citep[see also][]{xu07}.

\begin{figure}
\epsfxsize=76mm 
{\hfill 
\epsfbox{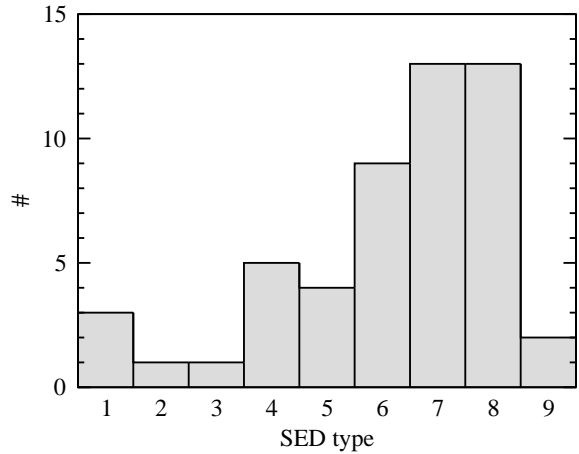} 
\hfill}
\caption{Histogram of best fit SED types for the 51 primary
SHADES sources listed in Table \ref{tab_sources}. SED templates are;
1=E, 2=S0, 3=Sa, 4=Sb, 5=Sc, 6=Sc+Im, 7=Im, 8=SB, 9=QSO.}
\label{sed_dist}
\end{figure}

Figure \ref{sed_dist} shows the distribution of SED type (listed in
Table \ref{tab_sources}) for the 51 primary sources. As the figure
shows, the SHADES sources are dominated by late type SEDs. Over 80 per
cent of sources have SEDs of type Sc or later and approximately two
thirds of these are best fit with either the Im or SB template. There
are only two sources that are best fit with a QSO template:
Lock850.075 and Lock850.081 (there is also the secondary
Lock850.041b).  This is entirely consistent with the low fraction of
AGN dominated SCUBA sources found in previous studies
\citep[e.g.,][]{ivison02,waskett03,almaini03,alexander03}

\begin{table*}
\caption{Photometric redshifts of the Lockman Hole SHADES
sources. Those sources with an ID suffix 'b' are the secondary
counterparts listed here for completeness but discounted from all
analysis (see Section \ref{sec_shades_cat}).  Codes in column labelled
'C' indicate whether the source has a radio and/or 24$\mu$m
counterpart. First character corresponds to radio and second to
24$\mu$m with 'R' indicating a robust identification, 'N' a non-robust
identification and '-' no detection (see Section
\ref{sec_shades_cat}).  All co-ordinates are either the radio or
24$\mu$m co-ordinates from \citet{ivison07} depending on which is the
more likely identification (exception is Lock850.036 -- see text).
Columns $z^+$ and $z^-$ give the 90 per cent uncertainty range on the
photometric redshift, $z$. Column `SED' lists best fit template number
(see text). Absolute AB magnitude listed under M$_K$.  $\chi^2$ of the
SED fit and the fractional area contained under the peak of the
probability distribution for $\chi^2$, p($z$), are given in columns 11
and 12 (see text).  Columns 13 to 15 list secondary redshift solutions
and their corresponding $\chi^2$ and p. Last column gives
spectroscopic redshifts from either (1) \citet{ivison05} or (2)
\citet{chapman05} (robust redshifts in bold). }
\centering
\begin{tabular}{@{}lcllcccccccccccc@{}}
\hline
ID & C & RA & Dec & $z$ & $z^-$ & $z^+$ & SED & $A_V$ & M$_K$ & $\chi^2(z)$ & p($z$) & $z2$ & $\chi^2(z2)$ & p$(z2)$ & $z_{\rm spec}$\\
\hline
Lock850.001 & RR & 10 52 01.249 & +57 24 45.76 & 4.21 & 2.41 & 4.45 & 6 & 0.2 & -24.15 & 0.76 & 0.87 &  -   &   -  &  -   & 2.15$^1$\\
Lock850.002 & RR & 10 52 57.084 & +57 21 02.82 & 3.23 & 2.76 & 3.34 & 7 & 0.7 & -24.10 & 0.22 & 0.98 &  -   &   -  &  -   & -\\
Lock850.003 & RR & 10 52 38.299 & +57 24 35.76 & 1.21 & 1.12 & 1.25 & 4 & 0.0 & -22.91 & 0.69 & 1.00 &  -   &  -   &  -   & 3.04$^1$\\
Lock850.003b& RR & 10 52 38.401 & +57 24 39.50 & 1.51 & 1.24 & 1.61 & 2 & 0.1 & -23.36 & 0.99 & 0.36 & 0.71 & 1.02 & 0.37 & -\\
Lock850.004 & RR & 10 52 04.079 & +57 26 58.52 & 2.66 & 2.25 & 2.85 & 7 & 0.9 & -24.71 & 0.26 & 0.73 & 3.16 & 0.73 & 0.26 & 0.53$^1$\\
Lock850.004b& RN & 10 52 04.226 & +57 26 55.46 & 2.11 & 1.59 & 2.50 & 6 & 0.9 & -24.26 & 0.67 & 1.00 &  -   &  -   &  -   & {\bf 1.48$^1$}\\
Lock850.005 & -N & 10 53 02.696 & +57 18 21.95 & 1.20 & 0.87 & 1.24 & 2 & 0.1 & -22.96 & 0.52 & 0.96 &  -   &  -   &  -   & -\\
Lock850.006 & RR & 10 52 04.013 & +57 25 24.20 & 2.01 & 1.81 & 3.18 & 8 & 2.0 & -23.67 & 0.60 & 0.98 &  -   &  -   &  -   & -\\
Lock850.007 & RR & 10 53 00.956 & +57 25 52.06 &   -  &   -  &   -  & - &  -  &     -  &  -   &  -   &  -   &  -   &  -   & -\\
Lock850.008 & -R & 10 51 53.690 & +57 18 34.90 &   -  &   -  &   -  & - &  -  &     -  &  -   &  -   &  -   &  -   &  -   & -\\
Lock850.009 & RR & 10 52 15.636 & +57 25 04.26 & 1.62 & 1.54 & 1.87 & 8 & 2.1 & -23.88 & 0.75 & 0.97 &  -   &  -   &  -   & {\bf 1.85$^1$}\\
Lock850.009b& -R & 10 52 15.730 & +57 25 01.70 & 0.91 & 0.82 & 1.01 & 6 & 0.4 & -23.01 & 0.43 & 1.00 &  -   &  -   &  -   & -\\
Lock850.010 & R- & 10 52 48.992 & +57 32 56.26 &   -  &   -  &   -  & - &  -  &     -  &  -   &  -   &  -   &  -   &  -   & -\\
Lock850.010b& -N & 10 52 48.270 & +57 32 51.00 & 1.33 & 1.26 & 1.39 & 7 & 1.1 & -22.09 & 0.91 & 0.98 &  -   &  -   &  -   & -\\
Lock850.011 & -N & 10 51 29.160 & +57 24 06.80 & 1.33 & 1.23 & 1.42 & 7 & 1.6 & -21.99 & 0.32 & 0.64 & 1.75 & 0.58 &  0.36& -\\
Lock850.012 & RR & 10 52 27.579 & +57 25 12.46 & 2.03 & 1.83 & 2.33 & 7 & 1.5 & -24.07 & 0.75 & 0.93 & 2.77 & 1.56 &  0.07& 2.14$^1$\\
Lock850.013 & -N & 10 51 31.770 & +57 31 41.20 & 0.48 & 0.31 & 0.55 & 7 & 0.1 & -21.70 & 0.61 & 1.00 &  -   &   -  &  -   & -\\
Lock850.014 & NN & 10 52 30.717 & +57 22 09.56 & 2.76 & 2.47 & 3.08 & 6 & 0.1 & -25.17 & 0.45 & 0.81 & 1.97 & 0.89 &  0.19& {\bf 2.61$^1$}\\
Lock850.015 & RR & 10 53 19.271 & +57 21 08.45 & 2.53 & 2.08 & 2.94 & 5 & 0.0 & -24.24 & 1.45 & 0.99 &  -   &  -   &  -   & -\\
Lock850.015b& R- & 10 53 19.025 & +57 21 09.47 &   -  &   -  &   -  & - &  -  &     -  &  -   &   -  &  -   &   -  &  -   & -\\
Lock850.016 & RR & 10 51 51.690 & +57 26 36.09 & 1.25 & 1.05 & 1.38 & 5 & 1.2 & -24.26 & 0.65 & 0.88 & 0.71 & 1.31 & 0.12 & {\bf 1.15$^1$}\\
Lock850.017 & RR & 10 51 58.018 & +57 18 00.27 & 1.12 & 1.00 & 1.38 & 6 & 1.3 & -23.39 & 0.28 & 1.00 &  -   &  -   &  -   & {\bf 2.24$^1$}\\
Lock850.017b& -R & 10 51 58.480 & +57 18 01.20 & 0.36 & 0.30 & 0.44 & 8 & 2.1 & -17.91 & 3.55 & 0.69 & 0.15 & 3.98 & 0.29 & -\\
Lock850.018 & R- & 10 52 27.778 & +57 22 18.18 &  -   &  -   &  -   & - &  -  &     -  &  -   &   -  &  -   &  -   &  -   & 1.96$^1$\\
Lock850.019 & -R & 10 52 36.090 & +57 31 19.60 & 3.69 & 3.56 & 3.75 & 6 & 0.3 & -24.35 & 3.56 & 0.88 & 3.00 & 4.30 & 0.11 & -\\
Lock850.021 & -R & 10 52 56.790 & +57 30 37.90 & 0.94 & 0.43 & 3.24 & 4 & 4.9 & -21.13 & 0.08 & 0.90 & 0.22 & 0.77 & 0.07 & -\\
Lock850.022 & -R & 10 51 37.090 & +57 33 16.90 & 2.68 & 2.55 & 2.90 & 8 & 0.1 & -24.11 & 0.89 & 0.93 & 1.86 & 2.24 & 0.07 & -\\
Lock850.023 & -N & 10 52 14.976 & +57 31 53.62 & 0.57 & 0.53 & 0.68 & 1 & 0.3 & -23.28 & 1.67 & 1.00 &  -   &  -   &  -   & -\\
Lock850.024 & RR & 10 52 00.445 & +57 20 40.16 & 1.40 & 1.25 & 1.49 & 8 & 1.5 & -24.12 & 0.72 & 1.00 &  -   &  -   &  -   & -\\
Lock850.026 & RR & 10 52 40.698 & +57 23 09.96 & 2.86 & 1.84 & 3.19 & 7 & 1.0 & -23.43 & 1.11 & 0.98 &  -   &  -   &  -   & -\\
Lock850.027 & -N & 10 52 03.450 & +57 18 19.30 & 1.12 & 1.03 & 1.20 & 7 & 0.4 & -22.47 & 1.17 & 1.00 &  -   &  -   &  -   & -\\
Lock850.028 & NN & 10 52 57.667 & +57 30 58.71 & 1.14 & 1.08 & 1.20 & 8 & 2.1 & -23.29 & 0.35 & 1.00 &  -   &  -   &  -   & -\\
Lock850.029 & N- & 10 51 31.305 & +57 20 40.28 &  -   &  -   &  -   & - &  -  &     -  &  -   &  -   &  -   &   -  &   -  & -\\
Lock850.030 & RR & 10 52 07.490 & +57 19 04.01 & 2.06 & 1.85 & 2.56 & 6 & 0.2 & -23.01 & 0.74 & 0.82 & 0.52 & 1.47 & 0.17 & {\bf 2.69$^2$}\\
Lock850.031 & RR & 10 52 15.989 & +57 16 19.34 & 2.12 & 1.95 & 2.31 & 6 & 1.0 & -24.44 & 0.81 & 1.00 &   -  &   -  &  -   & -\\
Lock850.033 & RN & 10 51 55.470 & +57 23 12.77 & 2.58 & 1.98 & 2.98 & 6 & 0.3 & -23.35 & 0.38 & 0.55 & 1.08 & 0.83 & 0.33 & {\bf 2.66$^1$}\\
Lock850.034 & RN & 10 52 14.202 & +57 33 28.30 & 3.42 & 3.15 & 3.69 & 7 & 0.6 & -24.05 & 0.92 & 1.00 &   -  &   -  &  -   & -\\
Lock850.035 & N- & 10 52 46.655 & +57 20 52.54 &  -   &  -   &  -   & - &  -  &     -  &  -   &   -  &   -  &   -  &  -   & -\\
Lock850.035b& -N & 10 52 45.940 & +57 20 51.40 & 1.59 & 1.41 & 1.71 & 7 & 0.6 & -22.68 & 1.03 & 1.00 &   -  &   -  &  -   & -\\
Lock850.036 & - -& 10 52 09.335 & +57 18 06.78 &  -   &  -   &  -   & - &  -  &     -  &  -   &   -  &   -  &   -  &  -   & -\\
Lock850.037 & R- & 10 51 24.342 & +57 23 36.18 &  -   &  -   &  -   & - &  -  &     -  &  -   &   -  &   -  &   -  &  -   & -\\
Lock850.037b& NR & 10 51 24.595 & +57 23 31.08 & 1.51 & 1.46 & 1.66 & 8 & 1.7 & -23.61 & 1.01 & 1.00 &   -  &   -  &  -   & -\\
Lock850.038 & RR & 10 53 07.060 & +57 24 31.60 & 1.29 & 1.17 & 1.39 & 8 & 1.2 & -23.67 & 0.85 & 1.00 &   -  &   -  &  -   & -\\
Lock850.039 & - -& 10 52 25.505 & +57 16 08.54 &  -   &  -   &  -   & - &  -  &     -  &  -   &   -  &   -  &   -  &  -   & -\\
Lock850.040 & RN & 10 52 01.721 & +57 19 17.00 & 2.02 & 1.87 & 2.14 & 6 & 0.6 & -23.88 & 2.92 & 0.98 &   -  &   -  &  -   & -\\
Lock850.041 & RR & 10 51 59.760 & +57 24 24.94 & 1.01 & 0.91 & 1.11 & 8 & 1.0 & -24.13 & 2.02 & 1.00 &   -  &   -  &  -   & -\\
Lock850.041b& RR & 10 52 00.248 & +57 24 21.69 & 2.22 & 2.05 & 2.44 & 9 & 0.9 & -24.37 & 0.86 & 0.52 & 0.32 & 1.17 & 0.26 & -\\
Lock850.043 & NR & 10 52 56.561 & +57 23 52.80 & 1.72 & 1.67 & 1.91 & 7 & 1.4 & -23.49 & 0.99 & 1.00 &   -  &   -  &  -   & -\\
Lock850.043b& NR & 10 52 56.576 & +57 23 58.62 & 1.14 & 1.04 & 1.42 & 8 & 1.4 & -23.96 & 0.39 & 1.00 &   -  &   -  &  -   & -\\
Lock850.047 & -N & 10 52 34.850 & +57 25 04.60 & 0.37 & 0.33 & 0.42 & 1 & 0.0 & -20.21 & 4.49 & 0.91 & 3.97 & 4.80 & 0.09 & -\\
Lock850.048 & -R & 10 52 56.030 & +57 32 42.30 & 0.31 & 0.24 & 0.37 & 8 & 1.6 & -21.28 & 0.81 & 1.00 &   -  &   -  &  -   & -\\
Lock850.052 & RR & 10 52 45.808 & +57 31 19.86 & 1.24 & 1.15 & 1.30 & 4 & 0.4 & -24.00 & 0.66 & 0.57 & 0.96 & 0.96 & 0.43 & -\\
Lock850.052b& -R & 10 52 46.160 & +57 31 20.20 & 1.40 & 1.28 & 1.60 & 5 & 1.5 & -23.62 & 0.81 & 0.84 & 0.71 & 1.63 & 0.12 & -\\
Lock850.053 & -R & 10 52 40.290 & +57 19 24.40 & 1.55 & 1.45 & 1.74 & 7 & 0.5 & -23.19 & 1.22 & 1.00 &   -  &   -  &  -   & -\\
Lock850.060 & -N & 10 51 43.900 & +57 24 43.60 & 1.92 & 1.12 & 2.34 & 4 & 0.1 & -22.38 & 1.11 & 0.60 &  2.69& 1.37 & 0.16 & -\\
\hline
\end{tabular}
\label{tab_sources}
\end{table*}

\begin{table*}
\contcaption{}
\begin{tabular}{@{}lcllcccccccccccc@{}}
\hline
ID & C & RA & Dec & $z$ & $z^-$ & $z^+$ & SED & $A_V$ & M$_K$ & $\chi^2(z)$ & p($z$) & $z2$ & $\chi^2(z2)$ & p$(z2)$ & $z_{\rm spec}$\\
\hline
Lock850.063 & RR & 10 51 54.261 & +57 25 02.55 & 4.73 & 4.62 & 4.89 & 7 & 0.2 & -25.50 & 0.14 & 1.00 &   -  &  -   &  -   & -\\
Lock850.064 & NN & 10 52 52.320 & +57 32 33.00 & 1.19 & 1.11 & 1.36 & 8 & 1.1 & -22.45 & 0.29 & 1.00 &   -  &  -   &  -   & -\\
Lock850.066 & -N & 10 51 39.570 & +57 20 27.10 & 1.41 & 1.01 & 1.94 & 8 & 0.3 & -20.70 & 0.05 & 0.54 & 2.15 & 0.27 & 0.33 & -\\
Lock850.067 & -R & 10 52 08.870 & +57 23 56.30 & 1.46 & 1.33 & 1.80 & 1 & 0.0 & -21.89 & 0.71 & 1.00 &   -  &  -   &  -   & -\\
Lock850.070 & NN & 10 51 47.894 & +57 30 44.37 & 0.53 & 0.46 & 0.62 & 7 & 1.2 & -20.86 & 0.84 & 1.00 &   -  &  -   &  -   & -\\
Lock850.071 & RN & 10 52 19.086 & +57 18 57.87 & 1.91 & 1.76 & 2.07 & 6 & 0.3 & -23.18 & 1.44 & 1.00 &   -  &  -   &  -   & -\\
Lock850.073 & RR & 10 51 41.992 & +57 22 17.52 & 1.40 & 1.31 & 1.50 & 8 & 1.2 & -23.76 & 0.16 & 1.00 &   -  &  -   &  -   & -\\
Lock850.073b& R- & 10 51 41.705 & +57 22 20.10 & 1.64 & 1.45 & 1.79 & 6 & 0.1 & -23.33 & 1.22 & 1.00 &   -  &  -   &  -   & -\\
Lock850.075 & -N & 10 53 15.190 & +57 26 45.90 & 2.06 & 1.95 & 2.28 & 9 & 1.3 & -23.87 & 0.77 & 0.55 & 0.29 & 1.07 & 0.35 & -\\
Lock850.076 & RR & 10 51 49.101 & +57 28 40.28 & 0.37 & 0.27 & 0.42 & 3 & 0.0 & -22.39 & 1.55 & 1.00 &  -   &  -   &  -   & -\\
Lock850.077 & RR & 10 51 57.153 & +57 22 09.58 & 1.76 & 1.68 & 1.88 & 5 & 0.0 & -21.96 & 1.27 & 0.65 & 1.34 & 1.58 & 0.34 & -\\
Lock850.077b& RN & 10 51 57.665 & +57 22 12.35 & 1.02 & 0.99 & 1.13 & 5 & 1.2 & -23.37 & 6.08 & 1.00 &  -   &  -   &  -   & -\\
Lock850.078 & -N & 10 51 44.088 & +57 17 44.52 & 1.48 & 1.39 & 1.85 & 8 & 0.5 & -21.73 & 0.08 & 0.39 & 0.58 & 0.23 & 0.31 & -\\
Lock850.079 & NR & 10 51 52.594 & +57 21 24.43 & 2.29 & 2.13 & 2.43 & 4 & 0.0 & -24.58 & 4.33 & 1.00 &  -   &  -   &  -   & -\\
Lock850.081 & NN & 10 52 31.523 & +57 17 51.67 & 2.19 & 2.07 & 2.37 & 9 & 1.4 & -27.51 & 2.45 & 1.00 &  -   &  -   &  -   & -\\
Lock850.083 & -R & 10 53 07.17  & +57 28 40.0  & 0.22 & 0.19 & 0.26 & 5 & 0.0 & -21.31 & 2.28 & 0.93 &  -   &  -   &  -   & -\\
Lock850.087 & RR & 10 51 53.365 & +57 17 30.05 & 2.46 & 2.11 & 2.80 & 7 & 1.6 & -24.94 & 1.29 & 1.0  &  -   &  -   &  -   & -\\
Lock850.100 & NN & 10 51 38.760 & +57 15 04.70 & 1.37 & 1.28 & 1.56 & 8 & 0.6 & -22.62 & 0.25 & 0.44 & 2.28 & 0.54 & 0.33 & -\\
\hline
\end{tabular}
\end{table*}

\section{Stellar Mass}
\label{sec_stellar_mass}

\subsection{Method of determination}
\label{sec_stellar_mass_method}

To determine stellar masses of the SHADES sources, we use the
synthetic spectra of \citet{bruzual03}. Rather than assume an a priori
model for the star formation history (SFH) of each galaxy, we divide
the history of each galaxy into blocks of fixed fractional time (see
Section \ref{sec_derived_mass}) and determine the amount of star
formation in each block. We choose five blocks as a compromise; too
few gives an unacceptably low SFH resolution whereas too many
introduces high degeneracy between blocks and reduces the number of
sources for which a stellar mass can be computed due to missing
photometry. Blocks are approximately logarithmically sized to account
for the fact that a galaxy's SED is more strongly influenced by more
recent star formation activity.

We use the 1994 Padova stellar evolutionary tracks \citep{bertelli94}
with a Salpeter initial mass function \citep{salpeter55}.  Starting
with a simple stellar population (SSP) SED, $L_{\lambda}^{\rm SSP}$,
we generate a composite stellar population (CSP) SED, $L_{\lambda}^{\, i}$,
for the $i$th block of constant star formation in a given galaxy using
\be
L_{\lambda}^{\, i} = \frac{1}{\Delta t_i} \int^{t_i}_{t_{i-1}} {\rm d}t' \, 
L_{\lambda}^{\rm SSP}(\tau(z)-t')
\ee
where the block lasts from time $t_{i-1}$ to $t_i$ in the galaxy's
history and $\tau$ is the age of the galaxy (i.e., the age of the
Universe today minus the lookback time to the galaxy). The SSP SED is
normalised to one solar mass hence the constant $\Delta t_i=t_i-t_{i-1}$
ensures that the CSP is also normalised to one solar mass. 
We then redden the CSP SED of each block using the extinction $A_V$
via the relationship
\be
L_{\lambda}^{\, i} \rightarrow L_{\lambda}^{\, i}
10^{-0.4 k(\lambda)A_V/R_V}
\ee
where $k(\lambda)$ is given by the Calzetti law for starbursts 
\citep{calzetti00},
\be
k(\lambda) = 
\left\{ \begin{array}{l}
2.659(-2.156+\frac{1.509}{\lambda}-\frac{0.198}{\lambda^2}+
\frac{0.011}{\lambda^3})+R_V \\
\hspace{15mm} ({\rm for}\,\,\, 0.12\mu{\rm m} < \lambda <  0.63\mu{\rm m}) \\
2.659(-1.857+\frac{1.04}{\lambda})+R_V \\
\hspace{15mm}  ({\rm for \,\,\,} 0.63\mu{\rm m} < \lambda <  2.2\mu{\rm m}) 
\end{array} \right .
\ee 
with $R_V=4.05$ and $\lambda$ in microns. To match our template
SED coverage, we assume that the longer wavelength half of the
function applies up to 10$\mu$m, and we linearly extrapolate the
shorter wavelength half down to 0.01$\mu$m using the average slope
between 0.12$\mu$m and 0.13$\mu$m\footnote{We have investigated the
effect of different extrapolations on our results and find negligible
dependence. Since the cut-off wavelength of 0.12$\mu$m does not reach
the central wavelength of our shortest filter ($B$ band) until $z\sim
3$, this is not surprising.}.

The flux (i.e., photon count) that would be observed in filter $j$ 
from a given block $i$ at the redshift $z$ of the galaxy is
\be
F_{ij}=\frac{1}{4\pi d_L^{\,2}}\int {\rm d}\lambda 
\frac{\lambda \, L^i_{\lambda}(\lambda/(1+z))T_j(\lambda)}{(1+z)\, hc} 
\ee
where $d_L$ is the luminosity distance and $T_j$ is the transmission
curve in filter $j$ (this includes telescope and atmospheric throughput
as well as detector response). We then find the normalisation $a_i$ 
of each block of star formation by minimising the $\chi^2$ function
\be
\label{eq_chi_sq}
\chi^2=\sum_j^{N_{\rm filt}} \frac{(\sum_i^{N_{\rm block}} \,
a_i F_{ij} - F^{\rm obs}_j)^2}{\sigma_j^2}
\ee
where $F^{\rm obs}_j$ is the flux observed in filter $j$ from the
galaxy and $\sigma_j$ is its error.  We treat non-detections
in the same way as we configured {\sl HyperZ} to, i.e., the
flux is set to zero and assigned a $1\sigma$ error equal to the
sensitivity of the corresponding filter.

We use a downhill simplex method to minimise $\chi^2$. To
prevent non-physical solutions we impose the constraint $a_i > 0;
i=1,N_{\rm block}$.  Since the CSP SED from which $F_{ij}$ is
computed is normalised to one solar mass, the quantity $a_i$ 
is the amount of stellar mass in solar units formed by the
galaxy in the time interval $\Delta t_i$. The total stellar mass of 
the galaxy is therefore simply
\be
{\rm M}_* = \sum_i^{N_{\rm block}} a_i \, .
\ee
The error on ${\rm M}_*$ is calculated by summing in quadrature 
the error on each $a_i$ derived from the fit.

In minimising $\chi^2$ in equation (\ref{eq_chi_sq}), we have the
choice of either minimising only the quantities $a_i$ and holding
$A_V$ and $z$ fixed at the values determined by {\sl HyperZ}, or
minimising all parameters $a_i$, $A_V$ and $z$. Although the latter
option seems the more compelling given its self-consistency, as we
show in Figure \ref{hyperz_vs_my_z}, the redshifts obtained in both
cases show very good agreement on the whole (see Section
\ref{sec_photz_results}).  Nevertheless, there are some significant
differences and since the templates used with {\sl HyperZ} are
empirical and include dust features, we choose the former option for
all analysis in this paper.

\subsection{Stellar mass results}
\label{sec_derived_mass}

We determined the stellar mass of each SHADES source using the
prescription given in the preceding section. An assumption of this is
that the metallicity is fixed as the source evolves.  Also, the same
metallicity is used for all sources.  To account for these
limitations, we treated the metallicity as a source of uncertainty,
repeating the calculation of stellar mass for two extremes.  To set
this range, we used the results of \citet{swinbank04} who measured the
metallicity of 30 high redshift SCUBA sources. They found an
approximately solar metallicity on average with a $1 \sigma$ scatter
of $\sim 0.25$dex.  We therefore performed two calculations of stellar
mass for each source, one using the spectra of \citet{bruzual03} with
metallicity Z = 0.4 Z$_\odot$ and a second with Z = 2.5 Z$_\odot$. The
resulting difference in mass was taken as the 68 per cent confidence
range and combined with the Monte Carlo error as described below.

\begin{figure}
\epsfxsize=80mm
{\hfill
\epsfbox{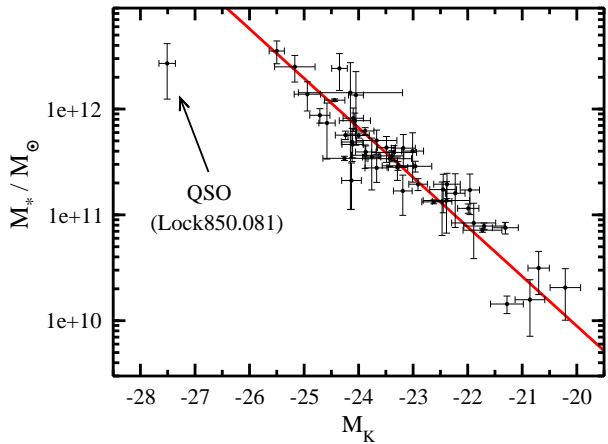}
\hfill}
\caption{Variation of stellar mass of the SHADES sources with
rest-frame absolute $K$ band magnitude (AB).  Mass error bars are
determined by a Monte Carlo analysis and include uncertainty due to
metallicity (see text) and errors on $M_K$ derived from redshift and
photometric uncertainty. The outlying point corresponds to
Lock850.081, best fit with the QSO template. Excluding this source,
the straight line fit is $\log_{10}\,{\rm
M}_*=(0.59\pm0.05)-(0.468\pm0.004)M_K$.}
\label{mass_vs_K}
\end{figure}

Figure \ref{mass_vs_K} shows the stellar masses computed for the
SHADES sources, plotted against the rest-frame absolute $K$ band
magnitude.  The correlation is very strong with little scatter.
Of the two sources best fit with QSO templates, Lock850.081
is an outlier (as labelled) but Lock850.075 lies within the main
trend with a slightly higher than median stellar mass of
$3.9\times10^{11}$ M$_{\odot}$. The most plausible explanation
for this difference is that unlike Lock850.081, the photometry of
Lock850.075 is not dominated by central QSO emission. If we
omit Lock850.081, linear regression gives the following relationship
\be
\label{eq_mass_mag_reln}
\log_{10}\,{\rm M}_*=(0.59\pm0.05)-(0.468\pm0.004)M_K \, .  
\ee
The rest-frame $K$ band flux is dominated by old stars and therefore,
as has been appreciated for some time, provides a good mass estimator
for such populations.  The small scatter seen in Figure
\ref{mass_vs_K} shows that rest-frame $K$ also provides a good
calibrator for the total stellar mass in SCUBA galaxies. This is a
direct consequence of the fact that the average SCUBA galaxy appears
to have an early and a late block of star formation which makes
approximately equal amounts of old and new stars (see next section).

The average stellar mass of our sample is $10^{11.8\pm 0.1} {\rm
M}_{\odot}$. This compares favourably with an average of $10^{11.4\pm
0.4} {\rm M}_{\odot}$ from the study of 13 SCUBA galaxies by
\citet{borys05}. Similarly, \citet{swinbank06} estimated dynamical
masses of eight submm galaxies, finding an average of $10^{11.7\pm
0.3} {\rm M}_{\odot}$. In terms of mass-to-light (M/L), the straight
line fit expressed by equation (\ref{eq_mass_mag_reln}) gives ${\rm
M/L} \propto {\rm L}^{0.17\pm0.01}$, indicating that the mass
increases more rapidly than the luminosity.

\begin{figure}
\epsfxsize=80mm
{\hfill
\epsfbox{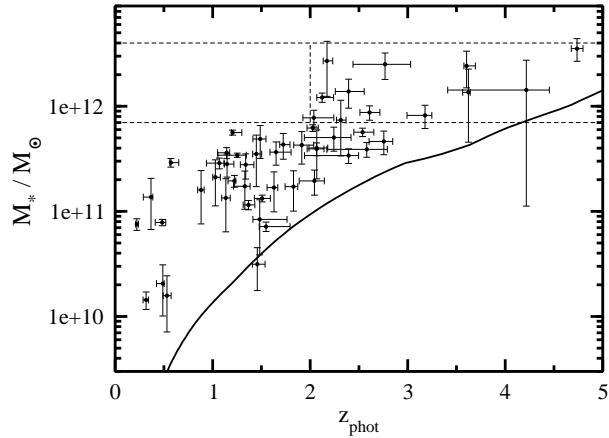}
\hfill}
\caption{Variation of stellar mass of the SHADES sources with redshift.
Mass error bars are determined by a Monte Carlo analysis and include
uncertainty due to metallicity (see text). The continuous curve shows
the mass limit transformed from the $K$ band sensitivity limit using
equation (\ref{eq_mass_mag_reln}) and our starburst SED template.  The
lack of objects within $7\times 10^{11}<{\rm M}_*/{\rm
M}_{\odot}<4\times 10^{12}$ at $z<2$ compared to the 12 objects within
$2<z<5$ is consistent with downsizing.}
\label{mass_vs_z}
\end{figure}

In Figure \ref{mass_vs_z}, we show how stellar mass varies with
photometric redshift. To obtain the mass error, we performed a Monte
Carlo analysis with 1000 realisations using solar metallicity SEDs. In
each realisation, we computed the mass, randomly sampling the fluxes
and redshift for each source using the measured errors and assuming a
normal distribution. The 1$\sigma$ scatter in stellar mass for each
source was then added in quadrature to the 1$\sigma$ error resulting
from the unknown metallicity as described above. These two errors are
typically approximately equal.

The continuous curve in Figure \ref{mass_vs_z} shows the mass
sensitivity limit. We estimate this by firstly computing the absolute
rest-frame $K$ band magnitude corresponding to the $5\sigma$ point
source sensitivity of 22.9 mag (AB) at each redshift assuming our
Im template SED.  This is then transformed into stellar mass
using the fitted relationship given in equation
(\ref{eq_mass_mag_reln}).

The horizontal dashed lines in Figure \ref{mass_vs_z} show the
arbitrary selection $7\times 10^{11}<{\rm M}_*/{\rm M}_{\odot}<4\times
10^{12}$. These limits select the 12 most massive sources. All 12
sources lie within $2<z<5$. Since the comoving volume over this
redshift interval is nearly twice that over $0<z<2$ (and since our
mass sensitivity barely affects the $2<z<5$ selection), one would
expect from simple Poisson statistics to find $\sim 6\pm2$ sources
within $0<z<2$. However, there are no SHADES sources with masses
greater than $7\times 10^{11}\,{\rm M}_{\odot}$ at $z<2$. This is
evidence in favour of downsizing, whereby star formation in the
Universe progressively shifts to smaller systems at later times
\citep[e.g.,][]{cowie96}.

\subsection{Star Formation Rate}
\label{sec_sfr}

In deriving these stellar masses, we divided the history of each
galaxy into the periods (0-0.45)$\tau$, (0.45-0.70)$\tau$,
(0.70-0.85)$\tau$, (0.85-0.95)$\tau$ and (0.95-1)$\tau$ where $\tau$
is the age of the galaxy, taken as the age of the Universe today minus
the lookback time to the source. All sources were therefore assumed to
start forming immediately after the Big Bang.

\begin{figure}
\epsfxsize=80mm
{\hfill
\epsfbox{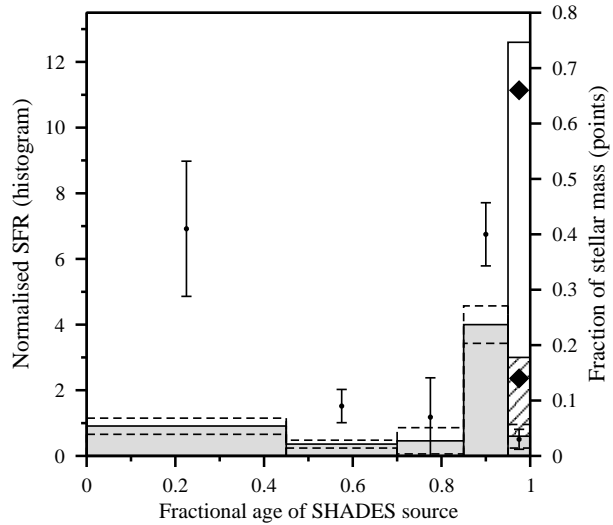}
\hfill}
\caption{The normalised SFR averaged over all 51 primary
SHADES sources (shaded histogram, left ordinate). Upper and lower
dashed histogram gives the 68 per cent confidence range allowing for
uncertainty in metallicity and including the Monte Carlo scatter
described in the text. The data points (right ordinate) give the
fraction of stellar mass created in each block on average.  The last
block also shows our estimate of the SFR and stellar mass created when
the star formation obscured at optical/IR wavelengths is taken into
account. The cross-hatched histogram and open histogram are computed
from the submm flux assuming a cold and hot SED respectively (see
Section \ref{sec_dust_correction}). The bold rhombi give the
corresponding stellar mass fractions.  In this last block, both the mass
fraction and SFR derived from the submm flux have been added to those
derived from the optical/IR.}
\label{av_sfh}
\end{figure}

The grey-shaded histogram in Figure \ref{av_sfh} shows the normalised
star formation rate\footnote{We define the normalised SFR as simply
the fractional stellar mass formed in each block divided by the block
width. Section \ref{sec_sfr_evol} considers the absolute SFR.}  (SFR)
in the five blocks, averaged over all 51 primary SHADES sources with
photometric redshifts. As in the determination of stellar mass, the
error, shown in the figure by the upper and lower dashed histograms,
incorporates the uncertainty due to the unknown metallicity and the
scatter from the Monte Carlo analysis. Clearly, the SFR is on average
dominated by a short burst close to the epoch at which the source is
observed. Since this result is derived from mainly rest-frame optical
photometry, the SFR in the last block will be suppressed due to
obscuration by dust. We estimate the effect of this in Section
\ref{sec_dust_correction}.

Determination of the average SFR in this way leads to a surprising
result. Even though the SFR is dominated by the later stages in the
average source's history, the {\em quantity} of stellar mass created,
i.e., the area under each block in Figure \ref{av_sfh}, is split
approximately evenly between these later stages and the earliest stage
(considering only the optical+IR derived results for now).
Specifically, during the earliest period, $(42\pm12)$ per cent of the
total stellar mass was formed, compared to $(39\pm6)$ per cent during
the fourth period. In the context of our model (see below), this
therefore indicates that the average SHADES source has already formed
a significant fraction of its stars \citep[also noted by][]{borys05}
and that it is undergoing a second major episode of star formation at
the epoch at which it is observed.

To verify the robustness of this result, we have carried out two
tests.  In the first test, we simply repeated the calculation twice
but using two different sub-sets of SHADES sources. In the first
sub-set, we selected those sources which meet the criteria
$\chi^2(z)\leq1.5$ and p$(z)\geq 0.9$. This leaves 27 out of the 51
primary sources with redshifts.  In the second sub-set, we selected
sources which have a robust identification in either the radio or at
24$\mu$m or both. This gives a sub-set of 35 sources. The repeat
calculation gives an almost identical result for both sub-sets, with
$\sim 40$ per cent of their stellar mass being created during the first
period and $\sim 40$ per cent being created in the fourth period.

In the second, more sophisticated test, we investigated how well our
method reconstructs an input SFH. We generated three synthetic
source catalogues each matching the number of SHADES sources and each
made with a different average SFH. We used the Bruzual and Charlot SED
library (with Salpeter IMF and Z = Z$_\odot$) to generate synthetic
photometry in our nine wavebands. For each source, we took the
redshift and $A_V$ determined by {\sl HyperZ} and assigned photometric
errors to the fluxes using a flux scaling relationship derived from
the real SHADES catalogue.  We then applied the analysis of Section
\ref{sec_stellar_mass_method} to assess how well the input average SFHs
could be recovered.

\begin{figure}
\epsfxsize=75mm
{\hfill
\epsfbox{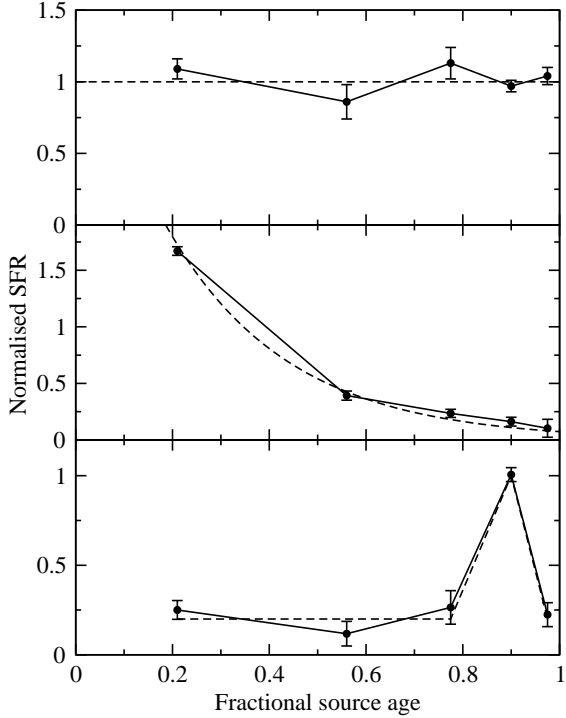}
\hfill}
\caption{Input (dashed line) and reconstructed (continuous line)
average SFHs. {\em Top}: Constant SFH. {\em Middle}: Exponential
decay SFH. {\em Bottom}: SFH matching the average exhibited by
the SHADES sources. The error bars include the Monte Carlo scatter
due to flux and redshift uncertainty.}
\label{simulated_sfh}
\end{figure}

Figure \ref{simulated_sfh} shows the results. The three panels from
top to bottom respectively show a constant SFH, a SFH undergoing
exponential decay and a SFH mimicking that exhibited by the SHADES
sources. The error bars in this plot include the Monte Carlo scatter
due to flux and redshift uncertainty as before.  The errors are smaller
than in Figure \ref{av_sfh} since each source is assigned exactly the
same SFH, unlike the real source population which will inevitably have
an intrinsic scatter. In all three cases, the reconstructed SFH very
faithfully reproduces the input SFH, indicating that the trend seen in
Figure \ref{av_sfh} is not an artifact of our method.

\subsubsection{SFR Evolution}

\begin{figure}
\epsfxsize=75mm
{\hfill
\epsfbox{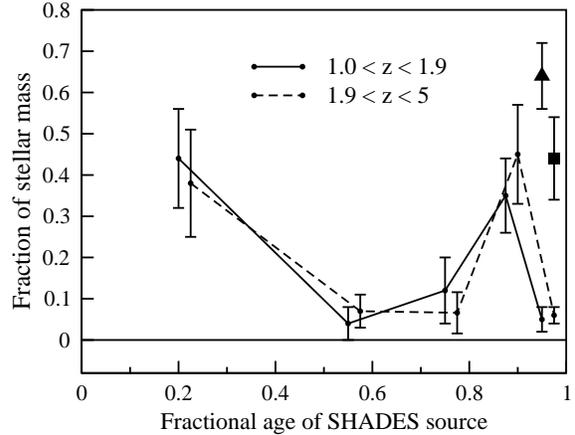}
\hfill}
\caption{Average fraction of stellar mass formed as a fraction of
SHADES source age for two different redshift bins ($\sim 22$ objects
per bin). Data points for the lower redshift bin are offset
horizontally by $-0.025$ for clarity. Errors account for uncertainty
in metallicity and include the Monte Carlo scatter described in the
text. The triangular and square points correspond respectively to the
low and high redshift slices and include an estimate of the stars
formed that are hidden by dust (see Section \ref{sec_dust_correction}
-- these points are averaged over the hot and cold SEDs).}
\label{av_sfh_evol}
\end{figure}

We have investigated the evolution of the trend seen in the SFR by
dividing the SHADES sources equally into a low and high redshift
bin. We defined the low redshift bin by $1< z \leq 1.9$ and the high
redshift bin by $1.9< z \leq 5$ (we limited this analysis to $z>1$
since the eight sources at $z<1$ poorly sample this large fraction of
the Universe's history). Figure \ref{av_sfh_evol} shows the SFHs for
the two redshift bins. We also show for the final SFH block (i.e.,
0.95-1 of the fractional age) our estimate of the star formation that
is completely obscured by dust (Section \ref{sec_dust_correction}). In
this case, for each redshift slice, we take the average of the SFR
computed from the hot and cold submm SED.

Figure \ref{av_sfh_evol} shows that there is little difference between
the fractional formation rate of stellar mass at high and low
redshift.  In both the high and low redshift sample, the initial and
late peaks in the rate of stellar mass formation persist. The late
peak for the low redshift sample is marginally less pronounced than
for the high redshift sample but still consistent within the errors.
Of course, in absolute terms, the late peak in the low redshift bin
corresponds to a very different star formation rate to that in the
high redshift bin for two reasons. Firstly, as Figure \ref{mass_vs_z}
shows, the median mass of a SHADES source in the high redshift bin is
$\sim 3$ times higher than the median mass in the low redshift
bin. Secondly, the proper time interval over which the late peak spans
is approximately twice that for a source at $z=1.5$ than for a source
at $z=3.5$. Therefore, on average, the late peak in the high redshift bin
corresponds to $\sim 5-10$ times the absolute star formation rate of
the peak in the low redshift bin.

\subsubsection{SFR: Dust correction}
\label{sec_dust_correction}

The results presented so far are based mainly on rest-frame optical
photometry. Although we have made a correction for extinction, in any
system undergoing massive star formation, some of the stars are
completely obscured at optical wavelengths. This is a particular
problem for the SHADES sources because we observe them as a result of
a massive starburst. In this section, we use the observed submm fluxes
to estimate the amount of hidden star formation.

We assume that the far-IR/submm luminosity of a SHADES source is the
result of a starburst extending in time over the period of the last
SFH block and that this is completely hidden in the optical/near-IR
wavelength range. We first calculate a pair of bolometric luminosities
for each source, $L^i_{\rm hot}$ and $L^i_{\rm cold}$, using the submm
flux and a hot and cold SED. These SEDs are the hottest and coolest
IRAS galaxy SEDs from the sample of
\citet{dunne01}\footnote{We choose to estimate the range of submm
bolometric flux in this way for homogeneity, rather than use
additional observational constraints (such as radio fluxes). The
coolest SED is that of NGC958 dominated by cold dust at 20K and the
hottest is IR1525+36 with a mix of dust at 26K and 57K in the ratio
15:1.}. We then calculate the hidden star-formation rate by taking the
ratio of this luminosity and the bolometric luminosity $L^i_{\rm
optical}$ for each source $i$ of the CSP for the final block generated
during the analysis of the previous section:
\be
{\rm SFR}=\sum_i L^i_{\rm [hot, cold]} \, \, /
\sum_i L^i_{\rm optical} \, .
\ee
This gives an upper or lower limit depending on whether the hot or cold
SED is used. Since the error caused by the uncertainty in metallicity
is insignificant compared to the range of SFRs spanned by the hot and 
cold SEDs, we ignore its contribution in this case.

The last block in Figure \ref{av_sfh} shows the additional hidden SFR
implied by the submm flux. The cross-hatched histogram bar is the
hidden SFR assuming the cold SED and the empty histogram bar
corresponds to the hot SED. Over the last 5 per cent of its history,
the hidden SFR of the average SHADES source is somewhere between 6 and
30 times the SFR implied by its optical flux. Similarly, the bold
rhombi show the range in the extra amount of stellar mass created in
this hidden burst which is 15-65 per cent of the total stellar mass
(Note, in the figure, both the stellar mass fraction and SFR estimated
from the submm have been added to the normalised quantities derived
from the optical+IR.)  If we therefore include the hidden star
formation, somewhere between 50 per cent and 65 per cent of the
stellar mass was created in the last 15 per cent of the average SHADES
source's history.

\subsection{The relationship of SCUBA sources to the field}

Figure \ref{av_sfh} shows that the average SHADES source is observed
during the most intense period of star formation the source has ever
experienced. An interesting question is whether all galaxies endure
such a phase, or whether the SHADES sources are rare in this
respect. Based on the most recent two blocks in the average SHADES
source's SFH, we can make the very crude statement that if all
galaxies have star formation histories like the average SHADES galaxy
found in this study, we would expect to see somewhere between 5 per
cent and 15 per cent of all galaxies undergoing a phase of highly
energetic dusty star formation at any one epoch.

To test whether this is the case, we constructed a plot of the
apparent $I$ band magnitude versus redshift for all the galaxies
detected in our $I$ band image of the Lockman Hole. Redshifts of the
field galaxies were obtained in exactly the same way as the SHADES
sources using {\sl HyperZ} with the same template SEDs. We formed a
master $I$ band catalogue using {\sl SExtractor} then coincidence matched
sources at all other wavelengths using a radial tolerance of $1''$. We
rejected stars using the CLASS\_STAR parameter output by {\sl SExtractor},
only retaining objects with CLASS\_STAR$<0.95$ (leaving 87 per cent of
objects). Objects in the vicinity of highly saturated stars were also
excluded to avoid deblending problems and pixel bleeding
(predominantly in the optical) as were objects detected in less than
five wavebands for consistency with the SHADES sample. This leaves a
total of 17,000 objects detected with 5 or more pixels above a
threshold of 2$\sigma$ in the full 320 sq. arcmin SHADES area.

\begin{figure}
\epsfxsize=75mm
{\hfill
\epsfbox{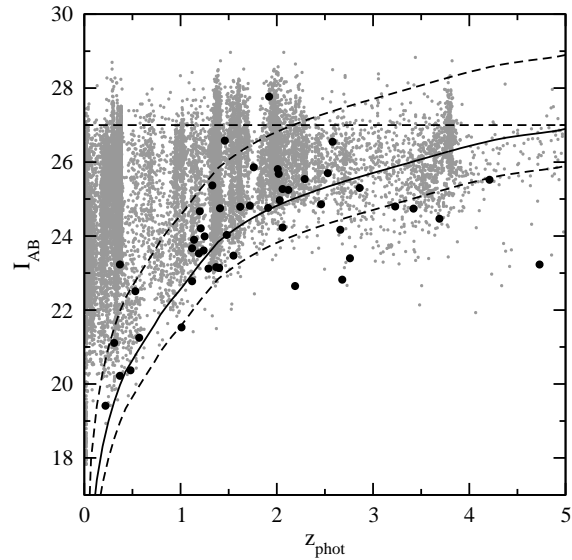}
\hfill}
\caption{Apparent $I$ band magnitude versus redshift for the 51
primary SHADES source counterparts listed in Table \ref{tab_sources}
(black points) compared to all galaxies in the field (grey
points).  The continuous curve corresponds to a non-evolving L$_*$ Im
galaxy. The dashed curves are the Im track offset by +2 mag and -1 mag
and the dashed straight line at $I_{\rm AB}=27.0$ is the 90 per cent
completeness limit.}
\label{I_vs_z}
\end{figure}

Figure \ref{I_vs_z} shows the $I-z$ plane for the SHADES sources
(large black points) and the full field galaxies (small grey
points). For comparison, the continuous curve shows the observed $I$
band flux of an L$_*$ galaxy computed assuming no evolution and using
our $I$ band filter response and Im template. The normalisation of
this curve is taken from \citet{blanton03} who measure M$_* = 5 \log h
- 20.82$ in $I$. The plot illustrates two important facts: 1) the
bright envelope of full field galaxies follows the L$_*$ galaxy track
fairly closely up to and beyond the median redshift, further
demonstrating that our photometric-redshifts are at least reasonable,
2) the average SHADES source is significantly brighter than the
average field galaxy.

Before the question regarding the rarity of SCUBA sources can be
addressed, we must make a few further considerations. The SHADES
galaxies have bright $I$ band magnitudes and are therefore among the
most luminous galaxies at every redshift. We wish to restrict our
comparison to field galaxies that are similarly luminous.  We
therefore limit both the SHADES and full field samples to those
objects that lie within -1 mag and +2 mag of the L$_*$ track (see
Figure \ref{I_vs_z}). Secondly, we apply an upper limit of $I_{\rm AB}
= 27$ where the 90 per cent completion limit takes effect. Finally, the
majority of the SHADES sources are at $z<3$ hence we limit our
computation of the fraction to three redshift bins, $0\leq z <1$,
$1\leq z <2$ and $2\leq z <3$.

\begin{table}
\centering
\begin{tabular}{@{}cc@{}}
\hline
Redshift & Fraction \\
\hline
$0\leq z <1$ & $(1.24\pm0.51)$ per cent \\
$1\leq z <2$ & $(0.72\pm0.17)$ per cent \\
$2\leq z <3$ & $(0.86\pm0.26)$ per cent \\
\hline
\end{tabular}
\caption{Evolution of the number of SHADES sources as a
fraction of galaxies in the full Lockman Hole field. Both the SHADES
sources and full field galaxies are selected by ${\rm L}_* +2{\rm mag}
< {\rm L} < {\rm L}_* -1{\rm mag}$ and $I_{\rm AB}\leq 27.0$. 
Errors assume Poisson noise only.}
\label{tab_fraction_evol}
\end{table}

Applying these constraints, we find in all three bins that
approximately 1 per cent of galaxies in the field is a SHADES source.
However, since we established from the average SFH that only 5-15 per
cent of SHADES sources would be seen during their most active period,
we can make the very approximate estimate that somewhere between one
in five and one in 15 bright galaxies in the field will experience a
highly energetic phase of dusty star formation.  This fraction shows
little evolution over $0\leq z < 3$ and is consistent within the
(large) errors with being constant at all redshifts in this range.

\subsection{SFR density evolution}
\label{sec_sfr_evol}

We determined the evolution of the SFR density using the 51 SHADES
sources for which we were able to obtain photometric redshifts. The
sources were divided into six equal redshift bins and the co-moving
volume density of the total SFR computed in each bin. For the SFR of
each galaxy, we took the average SFR in the last two SFH blocks with
and without the submm flux contribution.

\begin{figure*}
\epsfxsize=150mm
{\hfill
\epsfbox{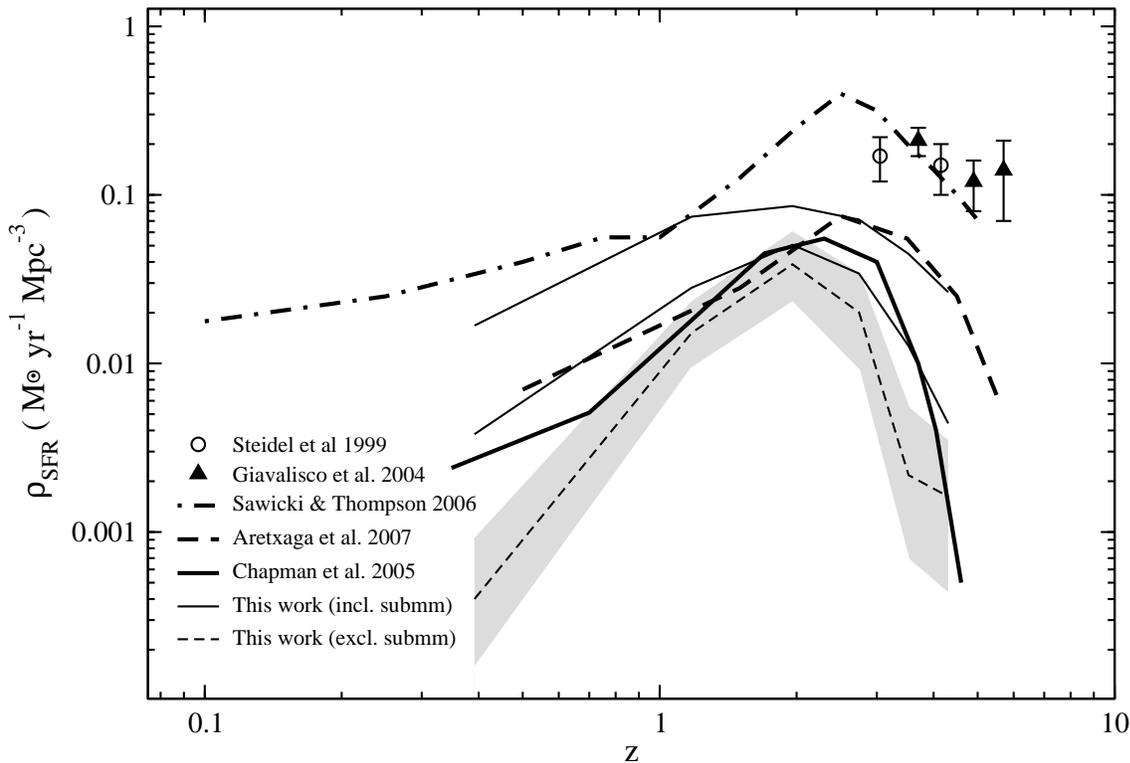}
\hfill}
\caption{The SFR density of the SHADES sources. The thin dashed line
is the mean density estimated from optical/near-IR photometry, with
the grey shading giving the 68 per cent confidence range. The thin
continuous lines show the same results but including the SFR implied
from the submm flux (upper for hot SED, lower for cool SED; see
Section \ref{sec_dust_correction}). Plotted for comparison are the
dust-corrected SFR densities derived from optical/UV observations of
\citet{sawicki06}, \citet{giavalisco04} and
\citet{steidel99}.  Also plotted are the SFRs determined from SCUBA
galaxies by \citet{chapman05} (extrapolated down to 1mJy) and
\citet{aretxaga07} (corrected by completing the luminosity function at
$60\mu$m).}
\label{rho_sfr}
\end{figure*}

The thin dashed line in Figure \ref{rho_sfr} shows our result from the
optical+IR photometry.  The 68 per cent confidence range is indicated
by the grey shading and accounts for the uncertainty in metallicity,
the Monte Carlo scatter as in previous sections and Poisson noise. The
thin continuous lines, in addition to the optical+IR photometry,
include the submm flux assuming the hot and cold submm SEDs described
in Section \ref{sec_dust_correction}. The peak in both cases lies
at $z \sim 2$, although the optical+IR peak is sharper than the
optical+IR+submm peak, implying that the submm flux is emitted over
a broader redshift interval. For comparison, we plot the 
SFR density measured from only submm flux by \citet{chapman05} and
\citet{aretxaga07}. Both of these show a peak around $2<z<3$, at a
slightly higher redshift than our peak but not significantly so.  The
height of our optical+IR+submm peak is also consistent with both
studies.

Compared to optical/UV observations such as the Keck Deep Field survey
of \citet{sawicki06} and the Lyman break galaxy surveys of
\citet{steidel99} and \citet{giavalisco04}, the SFR density is $\sim
5-10$ times lower for the SHADES sources. This suggests that at all
redshifts, most of the star formation is occurring in more modest
systems than the SHADES galaxies (which are at the bright end of the
luminosity function). However, since the surface number density of
SHADES sources in our sample is $\sim 100$ times lower out to a
redshift of $z\sim 4$ than the galaxies observed in these other
surveys, the average rate of star formation per SHADES source is $\sim
10-20$ times higher.

\section{Summary and Discussion}
\label{sec_summary}

Using nine-band photometry ranging from $B$ to 8$\mu$m, we have
determined best fit SEDs, photometric redshifts and stellar masses as
well as the average star formation history for 51 SCUBA sources in the
Lockman Hole. We find a median redshift of $z=1.52$ (with all objects
falling in the interval $0.22<z<4.73$), consistent with the median
photometric redshift found by \citet{clements07} for the SHADES SXDF
field.

Approximately 80 per cent of sources are best fit with late type
spectra ranging from Sc to starburst. Only two out of the 51 objects
are best fit with a QSO spectrum. Four objects are fit with an early
type (E or S0) SED, consistent with \cite{clements07} who find two
ellipticals in 33 sources. This is unexpected, especially since these
four sources all have low extinction ($A_V\leq
0.3$). Misidentifications aside, one possibility could be that these
objects have a very mixed stellar population. Young star forming
regions with a high column density of dust would be detected by SCUBA
whilst being heavily obscured in the rest-frame optical. Meanwhile,
old stellar regions with relatively little dust would be detected in
the rest-frame optical giving the appearance of an early type SED.

A surprising find is that the average SCUBA source has already built
up a significant fraction of its stellar mass in an early period of
star formation with the majority of the remainder being created in a
much later and more intense burst. This is consistent with the
findings of \citet{borys05} who concluded that the average SCUBA
source already has a massive population in place by $z=2.2$.
Including 850$\mu$m photometry indicates that a further 15-65 per cent
of the total stellar mass is created in an ongoing burst of dust
obscured star formation. The most recent 5-15 per cent of the average
SCUBA source's history (220 to 660 Myr on average) shows the highest
rates of star formation ever experienced by the source. Coupled with
the fact that $\sim 1$ per cent of bright field galaxies selected by
${\rm L}_* +2{\rm mag}<{\rm L_{optical}}<{\rm L}_* -1{\rm mag}$ over
the redshift range $0<z<3$ appear to be SCUBA galaxies, we estimate
that between one in five and one in 15 of these galaxies will at some
point in their lifetime experience a similar energetic burst of dust
obscured star formation.

We find that the trend of an early and late formation of stellar mass
with little intermediate activity does not differ between high
($1.9<z<5$) and low ($1<z<1.9$) redshifts. This suggests that the
typical SCUBA source is a snapshot of a system at the same point in
its history undergoing the same transformation process. This
transformation is from a system with an already established, mature
stellar population to a system with at least as much stellar mass
again. This is an intriguing result and fits neatly with the
conclusion of \citet{bell04} that the average elliptical galaxy has
doubled its mass since a redshift of $z=1$.  In addition to the early
and late periods of stellar mass formation, it is possible that there
could also have been so called `dry mergers' between two systems
containing old stars but little gas.  In such an event, very few new
stars would be formed and this therefore would not manifest itself in
the star formation history at the point of the merger, but would
enhance the stellar mass inferred from the early source history.

There is a distinct lack of SHADES sources in the redshift interval
$0<z<2$ with stellar masses greater than ${\rm M}_*=7\times
10^{11}\,{\rm M}_{\odot}$, compared to 12 sources within $2<z<5$ above
the same mass limit. This is clear evidence in favour of a downsizing
scenario, where star formation shifts to progressively smaller systems
as the Universe ages. \citet{clements07} have found exactly this trend
in the SXDF.

Finally, we have determined the evolution of the star formation rate
density using optical, IR and submm photometry. The peak occurs in the
vicinity of $z\sim 2$, consistent with that determined from submm only
studies \citep{chapman05,aretxaga07} and that derived from optical/UV
photometry \citep[e.g.,][]{sawicki06}. Since our sample amounts to a
total of only 51 sources, we are limited by Poisson noise throughout
most of the work presented here. This is especially true at $z<1$
where we detect only eight sources. Future investigations with
substantially more sensitive instruments such as SCUBA-2 and Herschel
will vastly improve this shortfall.

\appendix
\section{SED plots and postage stamps}
\label{sec_app_seds}

Figure \ref{seds} plots the best fit template SEDs to the SHADES
sources with redshifts listed in Table \ref{tab_sources}.

Postage stamp images for all wavelengths are illustrated in Figure
\ref{stamps}. 

\begin{figure*}
\epsfxsize=160mm
{\hfill
\epsfbox{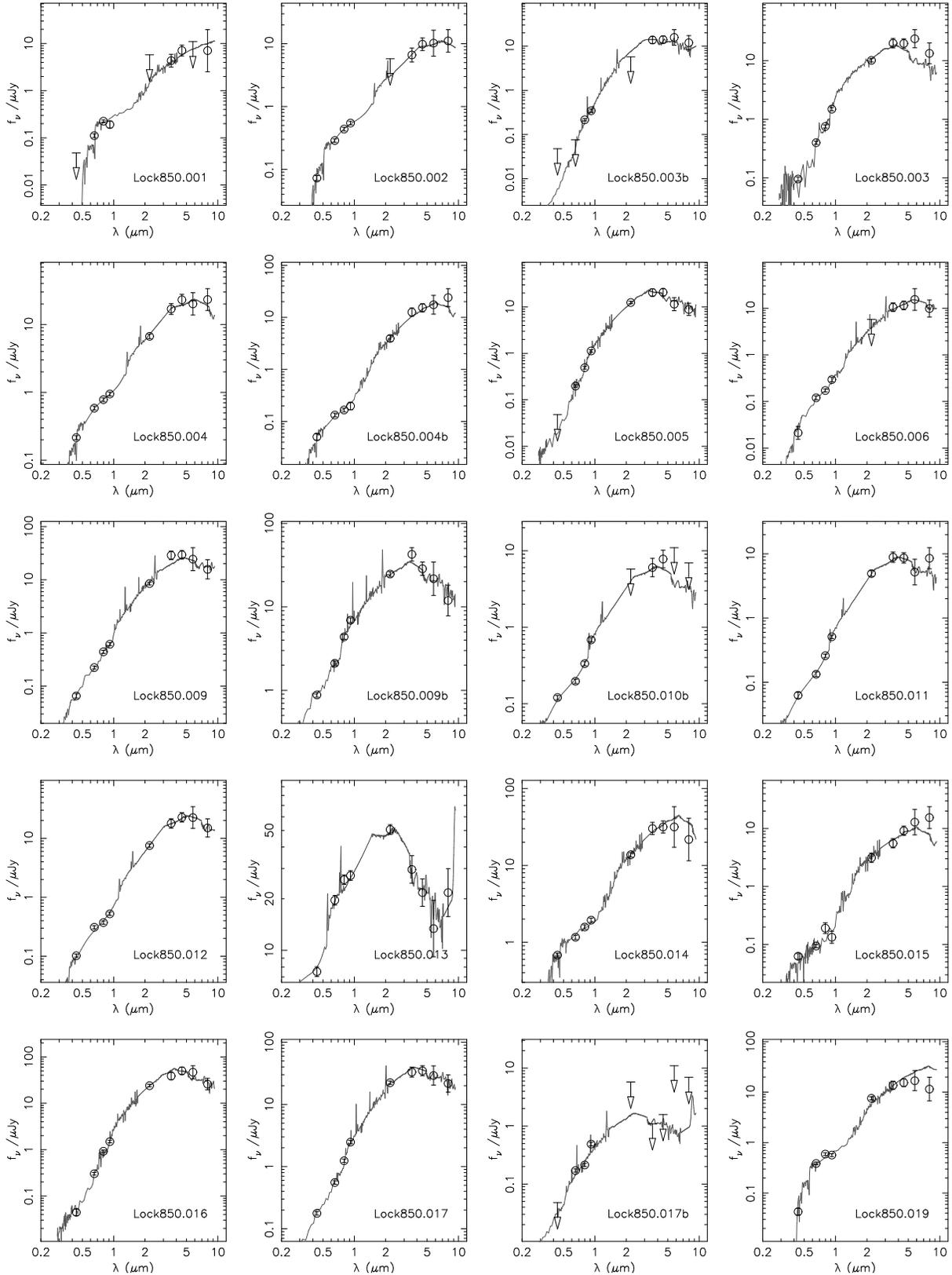}
\hfill}
\caption{Best fit SEDs for the SHADES sources determined by {\sl HyperZ}. 
Photometric points plot $1 \sigma$ error bars. Table \ref{tab_sources}
lists photometric redshifts, SED type, $A_V$ and absolute K band
magnitude for each source.}
\label{seds}
\end{figure*}

\begin{figure*}
\epsfxsize=160mm
{\hfill
\epsfbox{seds2.ps}
\hfill}
\contcaption{}
\end{figure*}

\begin{figure*}
\epsfxsize=160mm
{\hfill
\epsfbox{seds3.ps}
\hfill}
\contcaption{}
\end{figure*}

\begin{figure*}
\epsfxsize=160mm
{\hfill
\epsfbox{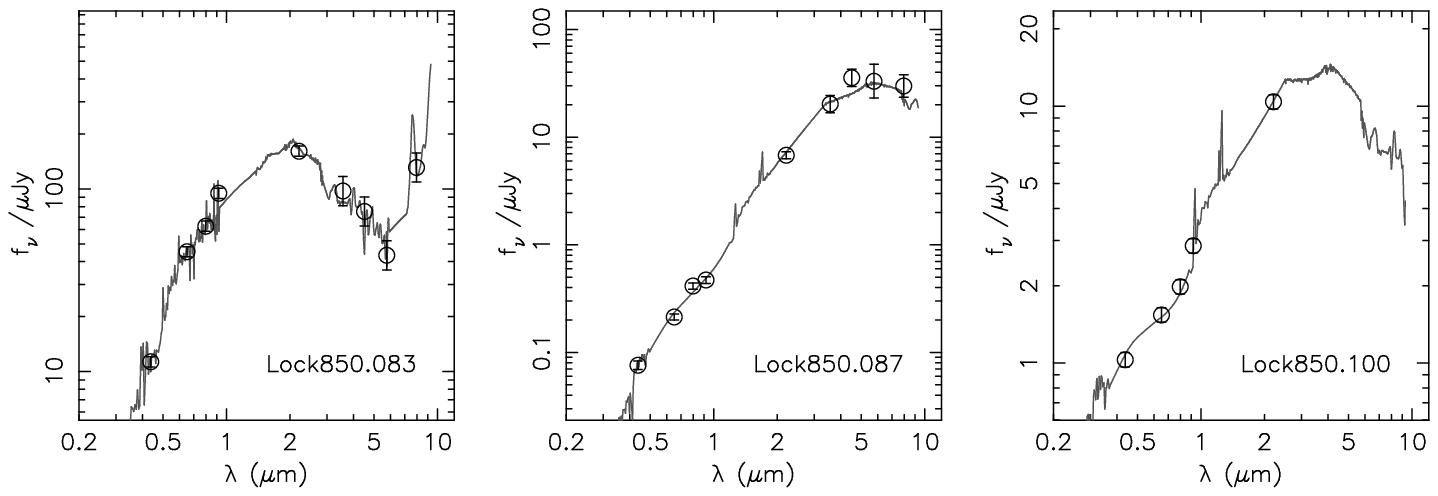}
\hfill}
\contcaption{}
\end{figure*}

\begin{figure*}
\epsfxsize=160mm
{\hfill
\hfill}
\caption{{\bf FIGURES OMITTED DUE TO astro-ph UPLOAD LIMIT}
Postage stamp images ($24'' \times 24''$) for all 60 SHADES
sources. Columns from left to right correspond to; $B$, $R$, $I$, $z$,
$K$, 3.6$\mu$m, 4.5$\mu$m, 5.8$\mu$m and 8$\mu$m. The cross-hair shows
the location of the primary counterpart and the diamond, where
present, shows the secondary.  The $10''$ radius circle is centred on
the SCUBA position. }
\label{stamps}
\end{figure*}

\begin{figure*}
\epsfxsize=160mm
{\hfill
\hfill}
\contcaption{}
\end{figure*}

\begin{figure*}
\epsfxsize=160mm
{\hfill
\hfill}
\contcaption{}
\end{figure*}

\begin{figure*}
\epsfxsize=160mm
{\hfill
\hfill}
\contcaption{}
\end{figure*}

\begin{figure*}
\epsfxsize=160mm
{\hfill
\hfill}
\contcaption{}
\end{figure*}

\section{Notes on sources}
\label{sec_app_source_notes}

In this appendix, we provide descriptions for a selection of
noteworthy sources.

\begin{flushleft}

{\bf Lock850.001:} This is a distinct source detected in $RIz$ and all
IRAC bands apart from 5.8$\mu$m. There is an unambiguous detection at
24$\mu$m and 1.4GHz. The source is probably blended with a fainter
neighbour to the W in the 3.6 and 4.5$\mu$m data. The SED and
redshift are derived from deblended photometry but using blended
photometry makes little difference to the resulting SED and redshift
of $z=4.2$.

{\bf Lock850.003:} There are two highly likely counterparts to this
source, both with robust radio and 24$\mu$m detections. The primary is
the fainter but more probable in both the radio and at 24$\mu$m with
an offset of $\sim 1''$ from the SCUBA position compared to $\sim
3''$ of the secondary.  The two counterparts have consistent
photometric redshifts of $z=1.21_{-0.09}^{+0.04}$ and
$z=1.51_{-0.27}^{+0.10}$. These are inconsistent with the
spectroscopic redshift of $z=3.04$ deemed non-robust by
\citet{ivison05}.

{\bf Lock850.004:} There are two robust counterparts,
both with significant detections at all wavelengths, in particular at
24$\mu$m and 1.4GHz. The primary counterpart has the lowest $P$ in both
the radio and at 24$\mu$m. The spectroscopic redshift for the primary
is considered non-robust. A third source lying $\sim 9''$ to the NNW
is brighter than the primary in the radio and at 24$\mu$m and has a
photometric redshift of $z=3.0$ with an early type SED.

{\bf Lock850.005:} The most likely counterpart is a faint non-robust
$24\mu$m source lying $\sim 5''$ from the SCUBA position in a SSE
direction.  There are no radio counterparts detected for this source.

{\bf Lock850.007:} There is a single robust radio and 24$\mu$m
counterpart but all wavebands show strong blending with a bright
neighbour which cannot be reliably deblended. This source is therefore
omitted from the photometric redshift analysis.

{\bf Lock850.008:} Optical photometry of this source is affected by
pixel bleeding from a nearby bright star and at IRAC wavelengths, the
source is blended with a neighbour of comparable brightness.  The
source is therefore omitted from the photometric redshift analysis.

{\bf Lock850.009:} The primary counterpart is the only robust radio
source and the most likely robust 24$\mu$m source. The secondary
counterpart included in Table \ref{tab_sources} has a robust 24$\mu$m
detection ($P=0.043$) offset $3.8''$ from the SCUBA position
but is not detected in the radio.

{\bf Lock850.010:} The robust radio counterpart reported in
\citet{ivison07} is only detected in the first two IRAC bands, hence
no photometric redshift can be determined for this source.  The
secondary counterpart in Table \ref{tab_sources} is the most likely
24$\mu$m counterpart, is detected at all wavelengths (apart from the
radio), has an offset of $\sim 8''$ SSW from the SCUBA position and is
not robust with $P$=0.31.

{\bf Lock850.011:} There are no robust radio or 24$\mu$m counterparts
for this source. The most likely counterpart is not detected in the
radio but is detected at 24$\mu$m with a near-robust $P$=0.053.

{\bf Lock850.015:} This source has two robust radio counterparts, both
close to each other ($\sim 2''$) and close to the SCUBA position (both
offset by $\sim 2''$). The primary counterpart is brighter in the
radio than the secondary and is the only source of the two to be
detected at 24$\mu$m. The secondary is included in Table
\ref{tab_sources} by virtue of having a robust radio detection but is
not detected at any other wavelength and hence has an undetermined
photometric redshift.

{\bf Lock850.017:} At optical wavelengths, the primary counterpart to
this source has a morphology comprising a centrally bright compact
nucleus and a more diffuse tail extending to the SW. An explanation
for the discrepancy between the photometric and spectroscopic
redshifts may be that the tail is a separate background object; the
photometry is dominated by flux from the compact object whereas
emission lines seen in the spectra may originate from the diffuse
object in the background. The secondary counterpart with a robust
24$\mu$m detection listed in Table \ref{tab_sources} is offset $\sim
2''$ due E, is very faint at all wavelengths and is best fit with the
SB template but with an unacceptable $\chi^2$.

{\bf Lock850.021:} No radio counterparts are detected for this
source. The only robust 24$\mu$m detection is taken as the primary
counterpart but this is only detected in the four IRAC wavebands. The
only other 24$\mu$m source detected within $15''$ of the SCUBA
position is non-robust, has an offset of $\sim 6''$ SW, is brighter
than the primary at all wavelengths and has a photometric
redshift of $z=1.4$ with an Im type SED.

{\bf Lock850.023:} \citet{ivison07} report two non-robust 24$\mu$m
counterparts with approximately equal values of $P$. There are no
radio sources detected. The slightly more probable (numerically lower
$P$) 24$\mu$m source is taken as the primary counterpart and has a
very faint 24$\mu$m flux and an SED that corresponds to a $z\simeq
0.1$ elliptical (consistent with its optical morphology). The less
probable 24$\mu$m source has a stronger 24$\mu$m flux and photometry
consistent with a starburst at $z \simeq 2.7$.

{\bf Lock850.027:} The primary counterpart has no radio detection but
a non-robust faint 24$\mu$m detection ($106\pm15 \mu$Jy) offset $\sim
6''$ to the N from the SCUBA position, consistent with an Im galaxy at
$z=1.1$.  Another possible counterpart, not listed in Table
\ref{tab_sources} since it is offset $\sim 11''$ to the SE, has a
faint non-robust 1.4GHz detection and a brighter but still non-robust
24$\mu$m detection ($196\pm13 \mu$Jy). The optical morphology and
photometry of this source are consistent with an S0 galaxy
at $z\simeq 0.6$.

{\bf Lock850.029:} No counterpart is detected at the position of the
most likely radio source noted by \citet{ivison07} at any other
wavelength. Another source lies $\sim 8''$ NW from the SCUBA position
and is detected at 24$\mu$m with $P$=0.15. This source is detected with
high S/N at all wavelengths, particularly at IRAC wavelengths (but is
not detected in the radio) and has photometry consistent with a Sa
galaxy at $z=1.4$.

{\bf Lock850.030:} This counterpart has strong radio and 24$\mu$m flux
and lies $\sim 3''$ to the SW of the SCUBA position. It is faint but
unambiguous at almost all other wavelengths, giving rise to a
relatively uncertain photometric redshift, marginally consistent with
the measured spectroscopic redshift.

{\bf Lock850.035:} The most probable counterpart to this source
reported by \citet{ivison07} is a non-robust radio source with a
probability $P=0.065$ offset to the SW by $\sim 5''$ from the SCUBA
position.  This source is not detected at any other wavelength,
therefore we include a secondary counterpart in Table
\ref{tab_sources} as an alternative possibility, this object having
the most probable 24$\mu$m detection seen at other wavelengths. This
secondary counterpart is offset by $\sim 9''$ SW of the SCUBA
position, has irregular optical morphology and is well fit by an Im
SED at $z=1.6$.

{\bf Lock850.036:} No radio or 24$\mu$m counterpart detected for this
source. Directly E at an offset of $\sim 8''$ from the SCUBA position
lies an irregular object which shows bright emission at IRAC
wavelengths. Surprisingly, the photometry of this source is well fit
by an elliptical SED with a redshift of $z=0.9$.

{\bf Lock850.037:} The most probable counterpart reported by
\citet{ivison07} is detected only in the radio. Table
\ref{tab_sources} includes a secondary counterpart with a robust
24$\mu$m detection and a non-robust radio detection lying $\sim 6''$
SE from the SCUBA position.  The secondary is the only source
detected within $10''$ at 5.8 and 8$\mu$m and the brightest of two at
24$\mu$m.

{\bf Lock850.041:} There are two very likely counterparts, both with
robust detections at 24$\mu$m and in the radio. Both have a high S/N
at all wavelengths. The most likely counterpart has photometry
consistent with a $z\sim 1$ starburst, whereas the slightly less
favourable counterpart (Lock850.041b) is very well fit by a QSO
template. As explained in the text (see Section
\ref{sec_photz_results}), Lock850.041b is most likely a $z\simeq2.2$
QSO strongly magnified by a $z=0.7$ \citep{ivison05} elliptical lens
$2''$ to the SW.

{\bf Lock850.043:} There are two robust 24$\mu$m counterparts detected
at all wavelengths for this source.  The primary is the nearer of the
two with a lower $P$ value and offset W by $\sim 5''$ from the SCUBA
position. The counterpart has a significant but not quite robust radio
detection and a weaker 24$\mu$m flux. The secondary counterpart is
slightly blended in $B$ with two neighbours lying to the E, becoming
less blended at longer wavelengths where it becomes dominant.  The
secondary counterpart lies $\sim 8''$ NW of the SCUBA position, has a
stronger 24$\mu$m flux but a very similar photometric redshift of
$z=1.20$.

{\bf Lock850.047:} The counterpart for this source has a very
disturbed optical morphology with a long ($\sim 4''$) `tail' extending
NNE. The radio and 24$\mu$m position are both located at the same
point in the `tail'. The photometry was measured in an aperture centred
on the radio position and is probably contaminated by the brighter
more compact source from which the tail appears to emanate. This most
likely causes the poor SED fit and hence uncertain photometric
redshift of $z=0.4$.

{\bf Lock850.052:} Both primary and secondary counterparts have
consistent photometric redshifts ($z \simeq 1.2$) and both are very
well fit with similar SED types (Sb/Sc). The primary, offset $\sim
3''$ ESE from the SCUBA position, has a robust radio and 24$\mu$m
detection, whereas the secondary, offset $\sim 5''$ ESE has a
robust 24$\mu$m detection but no radio detection.

{\bf Lock850.064:} There are three bright sources detected at IRAC
wavelengths within $10''$ of the SCUBA position. The primary
counterpart is the only one with significant 24$\mu$m flux although it
is non-robust being offset by nearly $10''$ to the SSE. The source is
detected in the radio (also non-robust) and is mildly blended with a
neighbour of similar brightness at IRAC wavelengths.

{\bf Lock850.073:} The primary and secondary counterparts have
consistent photometric redshifts and are well fit by late type SEDs.
The primary shows an irregular optical morphology and has robust radio
and 24$\mu$m detections. The secondary has a similar offset to the
primary from the SCUBA position ($\sim 2.6''$), has a robust radio
detection but is not detected at 24$\mu$m.

{\bf Lock850.077:} This source has two possible counterparts.  The
primary gains its status by virtue of being close to the SCUBA
position ($\sim 1.5''$ to the E) thereby having robust radio and
24$\mu$m detections. The brighter secondary (three times the flux in
the radio and at 24$\mu$m) is offset approximately $6''$ to the ENE of
the SCUBA position and the optical morphology clearly shows two very
close ($< 1''$) objects of equal flux and size. The $3''$ aperture we
use completely encompasses both of these objects belonging to the
secondary and therefore inaccurate deblending may explain the very
poor SED fit.

{\bf Lock850.083:} The only plausible counterpart for this source is a
nearby spiral galaxy with a photometric redshift of $z=0.22$ offset
$\sim 8''$ W of the SCUBA position.  The source has a robust 24$\mu$m
detection but is not detected in the radio. This object is best fit
with a type Sc SED, perfectly consistent with its optical morphology.

{\bf Lock850.100:} There are two very close potential counterparts for
this source. The primary counterpart is the brighter of the two,
with a photometric redshift of $z=1.4$ and having photometry
consistent with a starburst galaxy. The fainter source
is best fit with an Im SED at $z=0.4$. The IRAC photometry is
omitted for this object as the primary counterpart cannot be
deblended with its neighbour due to the IRAC PSF.

\end{flushleft}

\section{Photometric Data}
\label{sec_app_photom_data}

Table \ref{tab_photom_data} lists the optical to mid-IR photometry for
all SHADES source counterparts discussed in this paper. All magnitudes
are in the AB system and were extracted using an effective $3''$
diameter aperture (see Section \ref{sec_data} for more details).
All photometric errors were combined in quadrature with an error of
0.05 to account for uncertainties in zero point.

\begin{table*}
\caption{The optical to mid-IR photometry for all SHADES source
counterparts. Magnitudes are in the AB system and correspond to an
effective $3''$ diameter aperture (see Section \ref{sec_data}).
A numerical value of 99 corresponds to a non-detection,
whereas a hyphen indicates that photometry could not be
extracted. Errors include the zero point uncertainty in
each waveband. }
\centering
\begin{tabular}{@{}lccccccccc@{}}
\hline
ID & $B$ & $R$ & $I$ & $z$ & $K$ & 3.6$\mu$m & 4.5$\mu$m & 5.8$\mu$m 
& 8$\mu$m \\
\hline
Lock850.001 &       99         & 26.29$\pm$ 0.10 & 25.52$\pm$ 0.07 & 25.70$\pm$ 0.18 &      99         & 22.31$\pm$ 0.34 & 21.77$\pm$  0.29 &        99        & 21.78$\pm$  1.12 \\
Lock850.002 &  26.75$\pm$ 0.10 & 25.26$\pm$ 0.06 & 24.80$\pm$ 0.06 & 24.56$\pm$ 0.07 &      99         & 21.85$\pm$ 0.28 & 21.42$\pm$  0.25 & 21.38$\pm$  0.52 & 21.29$\pm$  0.45 \\
Lock850.003 &  26.45$\pm$ 0.09 & 24.90$\pm$ 0.06 & 24.21$\pm$ 0.06 & 23.47$\pm$ 0.06 & 21.39$\pm$ 0.06 & 20.66$\pm$ 0.21 & 20.67$\pm$  0.20 & 20.47$\pm$  0.39 & 21.09$\pm$  0.44 \\
Lock850.003b&       99         &      99         & 25.56$\pm$ 0.07 & 25.05$\pm$ 0.10 &      99         & 21.05$\pm$ 0.19 & 21.03$\pm$  0.21 & 20.90$\pm$  0.44 & 21.20$\pm$  0.40 \\
Lock850.004 &  25.57$\pm$ 0.06 & 24.48$\pm$ 0.06 & 24.17$\pm$ 0.06 & 23.96$\pm$ 0.06 & 21.84$\pm$ 0.08 & 20.83$\pm$ 0.20 & 20.49$\pm$  0.19 & 20.64$\pm$  0.41 & 20.48$\pm$  0.40 \\
Lock850.004b&  27.14$\pm$ 0.15 & 26.09$\pm$ 0.09 & 25.85$\pm$ 0.09 & 25.66$\pm$ 0.18 & 22.41$\pm$ 0.15 & 21.15$\pm$ 0.19 & 20.94$\pm$  0.18 & 20.79$\pm$  0.45 & 20.45$\pm$  0.43 \\
Lock850.005 &       99         & 25.66$\pm$ 0.06 & 24.67$\pm$ 0.06 & 23.77$\pm$ 0.06 & 21.16$\pm$ 0.06 & 20.62$\pm$ 0.20 & 20.61$\pm$  0.21 & 21.25$\pm$  0.35 & 21.53$\pm$  0.33 \\
Lock850.006 &  28.08$\pm$ 0.34 & 26.20$\pm$ 0.10 & 25.81$\pm$ 0.08 & 25.22$\pm$ 0.12 &      99         & 21.35$\pm$ 0.23 & 21.25$\pm$  0.23 & 20.93$\pm$  0.58 & 21.42$\pm$  0.46 \\
Lock850.007 &         -        &        -        &        -        &        -        &        -        &        -        &         -        &         -        &         -        \\
Lock850.008 &         -        &        -        &        -        &        -        &        -        &        -        &         -        &         -        &         -        \\
Lock850.009 &  26.86$\pm$ 0.11 & 25.53$\pm$ 0.06 & 24.79$\pm$ 0.06 & 24.43$\pm$ 0.06 & 21.58$\pm$ 0.06 & 20.25$\pm$ 0.20 & 20.23$\pm$  0.22 & 20.43$\pm$  0.54 & 20.91$\pm$  0.45 \\
Lock850.009b&  24.04$\pm$ 0.06 & 23.09$\pm$ 0.06 & 22.30$\pm$ 0.06 & 21.80$\pm$ 0.06 & 20.42$\pm$ 0.06 & 19.83$\pm$ 0.19 & 20.26$\pm$  0.16 & 20.56$\pm$  0.50 & 21.21$\pm$  0.46 \\
Lock850.010 &       99         &      99         &      99         &      99         &      99         & 22.71$\pm$ 0.40 & 22.27$\pm$  0.35 &        99        &        99        \\
Lock850.010b&  26.20$\pm$ 0.06 & 25.67$\pm$ 0.06 & 25.09$\pm$ 0.06 & 24.31$\pm$ 0.06 &      99         & 21.95$\pm$ 0.30 & 21.67$\pm$  0.29 &        99        &        99        \\
Lock850.011 &  26.91$\pm$ 0.12 & 26.09$\pm$ 0.09 & 25.37$\pm$ 0.06 & 24.63$\pm$ 0.07 & 22.17$\pm$ 0.11 & 21.53$\pm$ 0.19 & 21.56$\pm$  0.18 & 22.11$\pm$  0.50 & 21.57$\pm$  0.41 \\
Lock850.012 &  26.38$\pm$ 0.07 & 25.17$\pm$ 0.06 & 24.97$\pm$ 0.06 & 24.60$\pm$ 0.07 & 21.71$\pm$ 0.07 & 20.77$\pm$ 0.20 & 20.51$\pm$  0.18 & 20.52$\pm$  0.46 & 20.96$\pm$  0.39 \\
Lock850.013 &  21.71$\pm$ 0.06 & 20.67$\pm$ 0.06 & 20.37$\pm$ 0.06 & 20.31$\pm$ 0.06 & 19.64$\pm$ 0.06 & 20.22$\pm$ 0.20 & 20.56$\pm$  0.16 & 21.08$\pm$  0.41 & 20.56$\pm$  0.35 \\
Lock850.014 &  24.32$\pm$ 0.06 & 23.74$\pm$ 0.06 & 23.40$\pm$ 0.06 & 23.18$\pm$ 0.06 & 21.06$\pm$ 0.06 & 20.20$\pm$ 0.19 & 20.15$\pm$  0.21 & 20.15$\pm$  0.66 & 20.56$\pm$  0.69 \\
Lock850.015 &  26.91$\pm$ 0.12 & 26.47$\pm$ 0.07 & 25.70$\pm$ 0.23 & 26.09$\pm$ 0.26 & 22.66$\pm$ 0.20 & 22.04$\pm$ 0.20 & 21.50$\pm$  0.20 & 21.12$\pm$  0.55 & 20.93$\pm$  0.47 \\
Lock850.015b&       99         &      99         &      99         &      99         &      99         & 21.47$\pm$ 0.25 & 20.98$\pm$  0.21 &        99        &        99        \\
Lock850.016 &  27.28$\pm$ 0.17 & 25.20$\pm$ 0.06 & 23.99$\pm$ 0.06 & 23.47$\pm$ 0.06 & 20.45$\pm$ 0.06 & 19.93$\pm$ 0.21 & 19.65$\pm$  0.22 & 19.73$\pm$  0.35 & 20.37$\pm$  0.31 \\
Lock850.017 &  25.78$\pm$ 0.06 & 24.54$\pm$ 0.06 & 23.67$\pm$ 0.06 & 22.92$\pm$ 0.06 & 20.53$\pm$ 0.06 & 20.10$\pm$ 0.20 & 20.05$\pm$  0.18 & 20.24$\pm$  0.39 & 20.56$\pm$  0.35 \\
Lock850.017b&       99         & 25.82$\pm$ 0.07 & 25.58$\pm$ 0.07 & 24.68$\pm$ 0.08 &      99         &      99         &        99        &        99        &        99        \\
Lock850.018 &       99         &      99         &      99         &      99         &      99         &      99         &        99        &        99        &        99        \\
Lock850.019 &  27.32$\pm$ 0.17 & 24.93$\pm$ 0.06 & 24.47$\pm$ 0.06 & 24.52$\pm$ 0.07 & 21.71$\pm$ 0.08 & 21.06$\pm$ 0.20 & 20.93$\pm$  0.19 & 20.83$\pm$  0.50 & 21.25$\pm$  0.59 \\
Lock850.021 &       99         &      99         &      99         &      99         &      99         & 22.13$\pm$ 0.33 & 22.13$\pm$  0.35 & 21.42$\pm$  0.70 & 22.37$\pm$  0.60 \\
Lock850.022 &  23.65$\pm$ 0.06 & 23.05$\pm$ 0.06 & 22.82$\pm$ 0.06 & 22.74$\pm$ 0.06 & 21.77$\pm$ 0.08 & 21.27$\pm$ 0.19 & 21.01$\pm$  0.16 & 20.80$\pm$  0.40 & 21.07$\pm$  0.31 \\
Lock850.023 &  25.47$\pm$ 0.06 & 22.38$\pm$ 0.06 & 21.25$\pm$ 0.06 & 20.70$\pm$ 0.06 & 19.06$\pm$ 0.06 & 18.89$\pm$ 0.20 & 19.51$\pm$  0.21 & 19.71$\pm$  0.31 & 20.75$\pm$  0.30 \\
Lock850.024 &  24.61$\pm$ 0.06 & 23.74$\pm$ 0.06 & 23.14$\pm$ 0.06 & 22.65$\pm$ 0.06 & 20.52$\pm$ 0.06 & 19.70$\pm$ 0.20 & 19.59$\pm$  0.18 & 19.99$\pm$  0.32 & 20.62$\pm$  0.38 \\
Lock850.026 &  26.95$\pm$ 0.12 & 25.45$\pm$ 0.12 & 25.30$\pm$ 0.11 & 24.73$\pm$ 0.10 & 22.70$\pm$ 0.19 & 21.68$\pm$ 0.27 & 21.37$\pm$  0.24 & 20.90$\pm$  0.62 & 21.83$\pm$  0.53 \\
Lock850.027 &  23.77$\pm$ 0.06 & 23.55$\pm$ 0.06 & 22.78$\pm$ 0.06 & 22.32$\pm$ 0.06 & 21.18$\pm$ 0.06 & 20.90$\pm$ 0.20 & 21.00$\pm$  0.19 &        99        & 21.69$\pm$  0.45 \\
Lock850.028 &  25.90$\pm$ 0.06 & 24.60$\pm$ 0.06 & 23.90$\pm$ 0.06 & 23.27$\pm$ 0.06 & 20.88$\pm$ 0.06 & 20.29$\pm$ 0.20 & 20.20$\pm$  0.17 & 20.35$\pm$  0.37 & 20.88$\pm$  0.36 \\
Lock850.029 &       99         &      99         &      99         &      99         &      99         &      99         &        99        &        99        &        99        \\
Lock850.030 &  25.89$\pm$ 0.06 & 25.62$\pm$ 0.10 & 25.27$\pm$ 0.10 & 24.97$\pm$ 0.08 &      99         & 21.81$\pm$ 0.25 & 21.61$\pm$  0.24 &        99        &        99        \\
Lock850.031 &  26.56$\pm$ 0.09 & 25.81$\pm$ 0.07 & 25.25$\pm$ 0.06 & 24.94$\pm$ 0.09 & 21.76$\pm$ 0.07 & 20.44$\pm$ 0.20 & 20.32$\pm$  0.20 & 20.20$\pm$  0.58 & 20.98$\pm$  0.50 \\
Lock850.033 &  26.91$\pm$ 0.20 & 26.08$\pm$ 0.20 & 26.22$\pm$ 0.20 & 25.93$\pm$ 0.20 &      99         & 22.62$\pm$ 0.33 & 22.12$\pm$  0.27 &        99        & 21.85$\pm$  0.68 \\
Lock850.034 &  26.78$\pm$ 0.11 & 24.91$\pm$ 0.06 & 24.74$\pm$ 0.06 & 24.35$\pm$ 0.06 &      99         & 21.70$\pm$ 0.28 & 21.45$\pm$  0.26 & 21.12$\pm$  0.64 & 21.41$\pm$  0.57 \\
Lock850.035 &       99         &      99         &      99         &      99         &      99         &      99         &        99        &        99        &        99        \\
Lock850.035b&  24.65$\pm$ 0.06 & 24.31$\pm$ 0.06 & 23.83$\pm$ 0.06 & 23.64$\pm$ 0.06 & 21.74$\pm$ 0.08 & 21.51$\pm$ 0.21 & 21.33$\pm$  0.20 &        99        & 21.79$\pm$  0.55 \\
Lock850.036 &        -         &       -         &       -         &       -         &       -         &       -         &        -         &        -         &        -         \\
Lock850.037 &       99         &      99         &      99         &      99         &      99         &      99         &        99        &        99        &        99        \\
Lock850.037b&  25.81$\pm$ 0.06 & 24.81$\pm$ 0.06 & 24.09$\pm$ 0.06 & 23.87$\pm$ 0.06 & 21.14$\pm$ 0.06 & 20.40$\pm$ 0.18 & 20.23$\pm$  0.22 & 20.63$\pm$  0.43 & 21.32$\pm$  0.33 \\
Lock850.038 &  24.28$\pm$ 0.06 & 23.63$\pm$ 0.06 & 23.12$\pm$ 0.06 & 22.49$\pm$ 0.06 & 20.82$\pm$ 0.06 & 20.28$\pm$ 0.20 & 20.02$\pm$  0.21 & 20.53$\pm$  0.33 & 20.71$\pm$  0.30 \\
Lock850.039 &        -         &       -         &       -         &       -         &       -         &       -         &        -         &        -         &        -         \\
Lock850.040 &  26.33$\pm$ 0.07 & 25.95$\pm$ 0.08 & 25.68$\pm$ 0.08 & 24.97$\pm$ 0.10 & 22.47$\pm$ 0.16 & 21.21$\pm$ 0.19 & 20.95$\pm$  0.23 & 20.69$\pm$  0.51 & 21.36$\pm$  0.45 \\
Lock850.041b&  24.42$\pm$ 0.06 & 23.95$\pm$ 0.06 & 23.53$\pm$ 0.06 & 22.99$\pm$ 0.06 &        -        & 21.25$\pm$ 0.10 & 21.05$\pm$  0.22 & 20.42$\pm$  0.39 & 19.66$\pm$  0.37 \\
Lock850.041 &  22.95$\pm$ 0.06 & 22.05$\pm$ 0.06 & 21.53$\pm$ 0.06 & 21.05$\pm$ 0.06 & 19.96$\pm$ 0.06 & 19.27$\pm$ 0.16 & 19.26$\pm$  0.19 & 19.53$\pm$  0.40 & 19.69$\pm$  0.34 \\
Lock850.043 &  26.21$\pm$ 0.06 & 25.18$\pm$ 0.06 & 24.82$\pm$ 0.06 & 24.48$\pm$ 0.06 & 21.60$\pm$ 0.07 & 21.19$\pm$ 0.20 & 21.07$\pm$  0.19 & 20.88$\pm$  0.53 & 21.52$\pm$  0.42 \\
Lock850.043b&  23.66$\pm$ 0.30 & 23.01$\pm$ 0.15 & 22.46$\pm$ 0.11 & 21.82$\pm$ 0.06 & 20.14$\pm$ 0.06 & 19.69$\pm$ 0.16 & 19.52$\pm$  0.18 & 19.78$\pm$  0.36 & 20.39$\pm$  0.31 \\
Lock850.047 &  26.51$\pm$ 0.08 & 23.84$\pm$ 0.06 & 23.23$\pm$ 0.06 & 22.80$\pm$ 0.06 & 21.95$\pm$ 0.10 & 21.45$\pm$ 0.18 & 21.65$\pm$  0.23 & 22.21$\pm$  0.58 &        99        \\
Lock850.048 &  23.00$\pm$ 0.06 & 21.47$\pm$ 0.06 & 21.11$\pm$ 0.06 & 20.62$\pm$ 0.06 & 19.63$\pm$ 0.06 & 19.73$\pm$ 0.18 & 20.07$\pm$  0.21 & 20.53$\pm$  0.39 & 19.41$\pm$  0.32 \\
Lock850.052 &  26.17$\pm$ 0.06 & 24.56$\pm$ 0.06 & 23.61$\pm$ 0.06 & 22.88$\pm$ 0.06 & 20.39$\pm$ 0.06 & 19.30$\pm$ 0.20 & 19.64$\pm$  0.19 & 20.25$\pm$  0.39 & 20.71$\pm$  0.30 \\
Lock850.052b&       99         & 26.88$\pm$ 0.18 & 25.33$\pm$ 0.06 & 24.90$\pm$ 0.09 & 21.31$\pm$ 0.06 & 19.92$\pm$ 0.19 & 20.10$\pm$  0.20 & 20.52$\pm$  0.36 & 20.96$\pm$  0.39 \\
Lock850.053 &  24.21$\pm$ 0.06 & 23.92$\pm$ 0.06 & 23.47$\pm$ 0.06 & 23.19$\pm$ 0.06 & 21.58$\pm$ 0.06 & 20.96$\pm$ 0.21 & 20.90$\pm$  0.18 & 21.09$\pm$  0.35 & 21.46$\pm$  0.40 \\
Lock850.060 &       99         & 27.24$\pm$ 0.24 & 27.77$\pm$ 0.49 & 25.99$\pm$ 0.24 & 23.44$\pm$ 0.35 & 22.42$\pm$ 0.37 & 22.04$\pm$  0.33 &        99        &        99        \\
Lock850.063 &       99         & 24.57$\pm$ 0.15 & 23.23$\pm$ 0.06 & 23.11$\pm$ 0.06 & 21.86$\pm$ 0.08 & 20.98$\pm$ 0.22 & 20.75$\pm$  0.20 & 20.51$\pm$  0.36 & 20.47$\pm$  0.38 \\
Lock850.064 &  24.67$\pm$ 0.06 & 24.05$\pm$ 0.06 & 23.53$\pm$ 0.06 & 22.97$\pm$ 0.06 & 21.38$\pm$ 0.06 & 21.16$\pm$ 0.21 & 21.13$\pm$  0.20 & 21.35$\pm$  0.43 & 21.51$\pm$  0.41 \\
Lock850.066 &  25.40$\pm$ 0.06 & 25.06$\pm$ 0.06 & 24.75$\pm$ 0.06 & 24.49$\pm$ 0.06 &      99         & 23.15$\pm$ 0.52 & 23.10$\pm$  0.55 &        99        &        99        \\
Lock850.067 &  28.41$\pm$ 0.30 & 28.40$\pm$ 0.50 & 26.58$\pm$ 0.14 & 25.64$\pm$ 0.14 &      99         & 22.00$\pm$ 0.31 & 21.73$\pm$  0.30 & 22.38$\pm$  0.77 & 22.32$\pm$  0.68 \\
\hline
\end{tabular}
\label{tab_photom_data}
\end{table*}

\begin{table*}
\contcaption{}
\begin{tabular}{@{}lccccccccc@{}}
\hline
ID & $B$ & $R$ & $I$ & $z$ & $K$ & 3.6$\mu$m & 4.5$\mu$m & 5.8$\mu$m 
& 8$\mu$m \\
\hline
Lock850.070 &  24.67$\pm$ 0.05 & 23.10$\pm$ 0.06 & 22.51$\pm$ 0.06 & 22.41$\pm$ 0.06 & 21.03$\pm$ 0.06 & 21.13$\pm$ 0.20 & 21.58$\pm$  0.28 & 21.28$\pm$  0.32 &        99        \\
Lock850.071 &  25.43$\pm$ 0.06 & 24.97$\pm$ 0.06 & 24.76$\pm$ 0.06 & 24.20$\pm$ 0.06 & 21.76$\pm$ 0.06 & 21.46$\pm$ 0.25 & 21.05$\pm$  0.21 & 21.30$\pm$  0.35 &        99        \\
Lock850.073 &  24.35$\pm$ 0.06 & 23.63$\pm$ 0.06 & 23.14$\pm$ 0.06 & 22.67$\pm$ 0.06 & 20.65$\pm$ 0.06 & 20.33$\pm$ 0.20 & 20.22$\pm$  0.18 & 20.36$\pm$  0.42 & 20.59$\pm$  0.36 \\
Lock850.073b&  24.94$\pm$ 0.06 & 24.43$\pm$ 0.06 & 24.07$\pm$ 0.06 & 23.61$\pm$ 0.06 & 21.79$\pm$ 0.08 &        -        & 20.73$\pm$  0.19 &         -        & 20.78$\pm$  0.43 \\
Lock850.075 &  25.57$\pm$ 0.07 & 24.76$\pm$ 0.06 & 24.23$\pm$ 0.06 & 23.78$\pm$ 0.06 & 22.13$\pm$ 0.11 & 21.21$\pm$ 0.21 & 21.00$\pm$  0.20 &         -        &         -         \\
Lock850.076 &  23.23$\pm$ 0.06 & 21.03$\pm$ 0.06 & 20.22$\pm$ 0.06 & 20.05$\pm$ 0.06 & 19.00$\pm$ 0.06 & 19.10$\pm$ 0.17 & 19.47$\pm$  0.20 & 19.53$\pm$  0.33 & 19.36$\pm$  0.28 \\
Lock850.077 &  27.02$\pm$ 0.13 & 26.50$\pm$ 0.13 & 25.86$\pm$ 0.09 & 24.81$\pm$ 0.08 &      99         & 22.23$\pm$ 0.34 & 22.23$\pm$  0.36 &        99        &        99        \\
Lock850.077b&  27.59$\pm$ 0.12 & 24.97$\pm$ 0.06 & 24.62$\pm$ 0.06 & 23.44$\pm$ 0.06 & 21.14$\pm$ 0.06 & 20.12$\pm$ 0.20 & 20.03$\pm$  0.16 &         -        &         -         \\
Lock850.078 &  24.84$\pm$ 0.02 & 24.39$\pm$ 0.06 & 24.03$\pm$ 0.06 & 23.84$\pm$ 0.06 & 22.39$\pm$ 0.14 &        -        &         -        &         -        &         -         \\
Lock850.079 &  26.39$\pm$ 0.07 & 26.60$\pm$ 0.14 & 25.54$\pm$ 0.07 & 25.05$\pm$ 0.10 & 22.06$\pm$ 0.12 & 21.04$\pm$ 0.20 & 20.79$\pm$  0.20 & 20.65$\pm$  0.51 & 20.52$\pm$  0.36 \\
Lock850.081 &  23.75$\pm$ 0.05 & 23.18$\pm$ 0.06 & 22.65$\pm$ 0.06 & 22.19$\pm$ 0.06 & 20.07$\pm$ 0.06 & 19.59$\pm$ 0.19 & 18.90$\pm$  0.18 & 18.02$\pm$  0.20 & 17.20$\pm$  0.19 \\
Lock850.083 &  21.27$\pm$ 0.06 & 19.76$\pm$ 0.06 & 19.41$\pm$ 0.06 & 18.96$\pm$ 0.06 & 18.38$\pm$ 0.06 & 18.93$\pm$ 0.18 & 19.21$\pm$  0.19 & 19.81$\pm$  0.21 & 18.61$\pm$  0.20 \\
Lock850.087 &  26.70$\pm$ 0.10 & 25.58$\pm$ 0.07 & 24.86$\pm$ 0.06 & 24.72$\pm$ 0.08 & 21.82$\pm$ 0.08 & 20.63$\pm$ 0.21 & 20.02$\pm$  0.21 & 20.10$\pm$  0.39 & 20.21$\pm$  0.26 \\
Lock850.100 &  23.87$\pm$ 0.06 & 23.43$\pm$ 0.06 & 23.16$\pm$ 0.06 & 22.76$\pm$ 0.06 & 21.36$\pm$ 0.06 &        -        &         -        &         -        &         -         \\
\hline
\end{tabular}
\end{table*}

\begin{flushleft}
{\bf Acknowledgements}
\end{flushleft}

SD is supported by the Particle Physics and Astronomy Research
Council. We thank Ian Smail and an anonymous referee for several
helpful suggestions which have improved this manuscript. The work
presented in this paper is based partly on data collected at Subaru
Telescope operated by the National Astronomical Observatory of Japan,
partly on data acquired by The United Kingdom Infrared Telescope
operated by the Joint Astronomy Centre on behalf of the U.K. Science
and Technologies Facilities Council and partly on observations made
with the Spitzer Space Telescope operated by the Jet Propulsion
Laboratory, California Institute of Technology under a contract with
NASA. 


{}

\label{lastpage}

\end{document}